\documentclass[twocolumn,english,aps,showpacs,nofootinbib]{revtex4}
\usepackage[T1]{fontenc}
\usepackage[latin9]{inputenc}
\usepackage{amsmath}
\usepackage{amssymb}
\usepackage{appendix}
\usepackage{esint}
\usepackage{graphicx}
\usepackage{relsize}

\makeatletter
\@ifundefined{textcolor}{}
{%
 \definecolor{BLACK}{gray}{0}
 \definecolor{WHITE}{gray}{1}
 \definecolor{RED}{rgb}{1,0,0}
 \definecolor{GREEN}{rgb}{0,1,0}
 \definecolor{BLUE}{rgb}{0,0,1}
 \definecolor{CYAN}{cmyk}{1,0,0,0}
 \definecolor{MAGENTA}{cmyk}{0,1,0,0}
 \definecolor{YELLOW}{cmyk}{0,0,1,0}
 }

\usepackage{nicefrac}\usepackage{dsfont}%\usepackage{xfrac}
\usepackage[normalem]{ulem}\@ifundefined{definecolor}
 {\usepackage{color}}{}

\makeatother

\usepackage{babel}
\begin{document}

\title{Strain-driven criticality underlies nonlinear mechanics of fibrous networks }
\author{A. Sharma\textsuperscript{1,2}, A. J. Licup\textsuperscript{1}, R. Rens\textsuperscript{1}, M. Vahabi\textsuperscript{1}, K. A. Jansen\textsuperscript{3,4},  G. H. Koenderink\textsuperscript{3}, F. C. MacKintosh\textsuperscript{1,5}}
\address{\textsuperscript{1}Department of Physics and Astronomy, VU University, Amsterdam, The Netherlands\\
\textsuperscript{2}Department of Physics, University of Fribourg, CH-1700 Fribourg, Switzerland\\
\textsuperscript{3}FOM Institute AMOLF, Science Park 104, 1098 XG Amsterdam, The Netherlands\\
\textsuperscript{4}Wellcome Trust Centre for Cell-Matrix Research, Faculty of Life Sciences, University of Manchester, Manchester M13 9PT, UK\\
\textsuperscript{5}Departments of Chemical and Biomolecular Engineering, Chemistry and Physics, Rice University, Houston, TX 77005, USA}

\date{\today}
\begin{abstract}
Networks with only central force interactions are floppy when their average connectivity is below an isostatic threshold. 
Although such networks are mechanically unstable, they can become rigid when strained. 
It was recently shown that the transition from floppy to rigid states as a function of simple shear strain is continuous, with hallmark signatures of criticality~\cite{sharma2016strain}. 
The nonlinear mechanical response of collagen networks was shown to be quantitatively described within the framework of such mechanical critical phenomenon.
Here, we provide a more quantitative characterization of critical behavior in subisostatic networks. 
Using finite size scaling we demonstrate the divergence of strain fluctuations in the network at well-defined critical strain. 
We show that the characteristic strain corresponding to the onset of strain stiffening is distinct from but related to this critical strain in a way that depends on critical exponents. We confirm this prediction experimentally for collagen networks. Moreover, we find that the apparent critical exponents are largely independent of the spatial dimensionality. In a highly simplified computational model of network dynamics, we also observe critical slowing down in the vicinity of the critical strain. With subisostaticity as the only required condition, strain-driven criticality is expected to be a general feature of biologically relevant fibrous networks. 
\end{abstract}

\maketitle
Disordered filamentous networks are ubiquitous in biology. An important example of such networks is the extracellular matrix of in biological tissues which is predominantly composed of a fibrous collagen scaffold~\cite{fratzl2008collagen}. One of the most important characteristics of such networks is the coordination number or average connectivity $\langle z \rangle$. Networks with only central force interactions are unstable towards small deformation if the average connectivity is below the threshold value of $\langle z \rangle = 2d$, where $d$ is the dimensionality. This threshold is referred to as the \emph{isostatic} point at which, as shown by Maxwell~\cite{maxwell1864}, the number of degrees of freedom are just balanced by the number of constraints, and the system is \emph{marginally stable}. As the average connectivity increases beyond $2d$, the network undergoes a phase transition marked by a continuous increase in the elasticity. Other examples of such transitions are the jamming transition~\cite{liu1998nonlinear,majmudar2007jamming,van2010jamming,saarloos2010jamming} in granular materials and rigidity percolation~\cite{thorpe1983continuous,feng1984percolation,jacobs1995generic,latva2001rigidity} in disordered spring networks. Jamming exhibits signatures characteristic of both first- and second-order transitions, with discontinuous behavior of the bulk modulus and continuous variation of the shear modulus~ \cite{olsson2007critical,head2009critical,van2010jamming}. For networks of springs or fibers, the transition from floppy to rigid is a continuous phase transition, in both bulk and shear moduli, with critical signatures~\cite{thorpe1983continuous,wyart2008elasticity,ellenbroek2009non,van2010jamming,chase2011,sheinman2012nonlinear}.

In a biological context, the average connectivity is almost always below the isostatic threshold. Filamentous networks typically fall in two categories, those in which network formation occurs via branching and those where crosslinking proteins connect two distinct filaments. The typical connectivity in such networks is between 3 and 4, with the former due to branching and the latter due to binary crosslinking. In fact these networks are well below both 2D and 3D isostatic thresholds~\cite{chase2011,licup2015elastic}. Such subisostatic networks can, however, become rigid as a result of other mechanical constraints, such as fiber bending~\cite{Head2003PRL,wilhelm2003elasticity,wyart2008elasticity,chase2011}, internal stresses~\cite{sheinman2012actively}, thermal fluctuations~\cite{dennison2013fluctuation}, or when subjected to external strain~\cite{alexander1998amorphous,sheinman2012nonlinear}. Except for the external strain, other applied fields stabilize the network even in the zero strain limit, i.e., the subisostatic network becomes stable to small deformations. However, when the applied field is an external strain, the transition from floppy to rigid states occurs at a threshold strain which depends on the network structure, nature of the applied deformation as well as the average connectivity~\cite{sheinman2012nonlinear}. We recently showed that sheared subisostatic networks exhibit a line of second order transitions at a strain threshold $\gamma_c(z)$, for connectivities $\langle z\rangle$ well below the isostatic threshold~\cite{sharma2016strain}. 

\begin{figure*}
\begin{tabular}{@{}c@{}}
\includegraphics[width = \columnwidth]{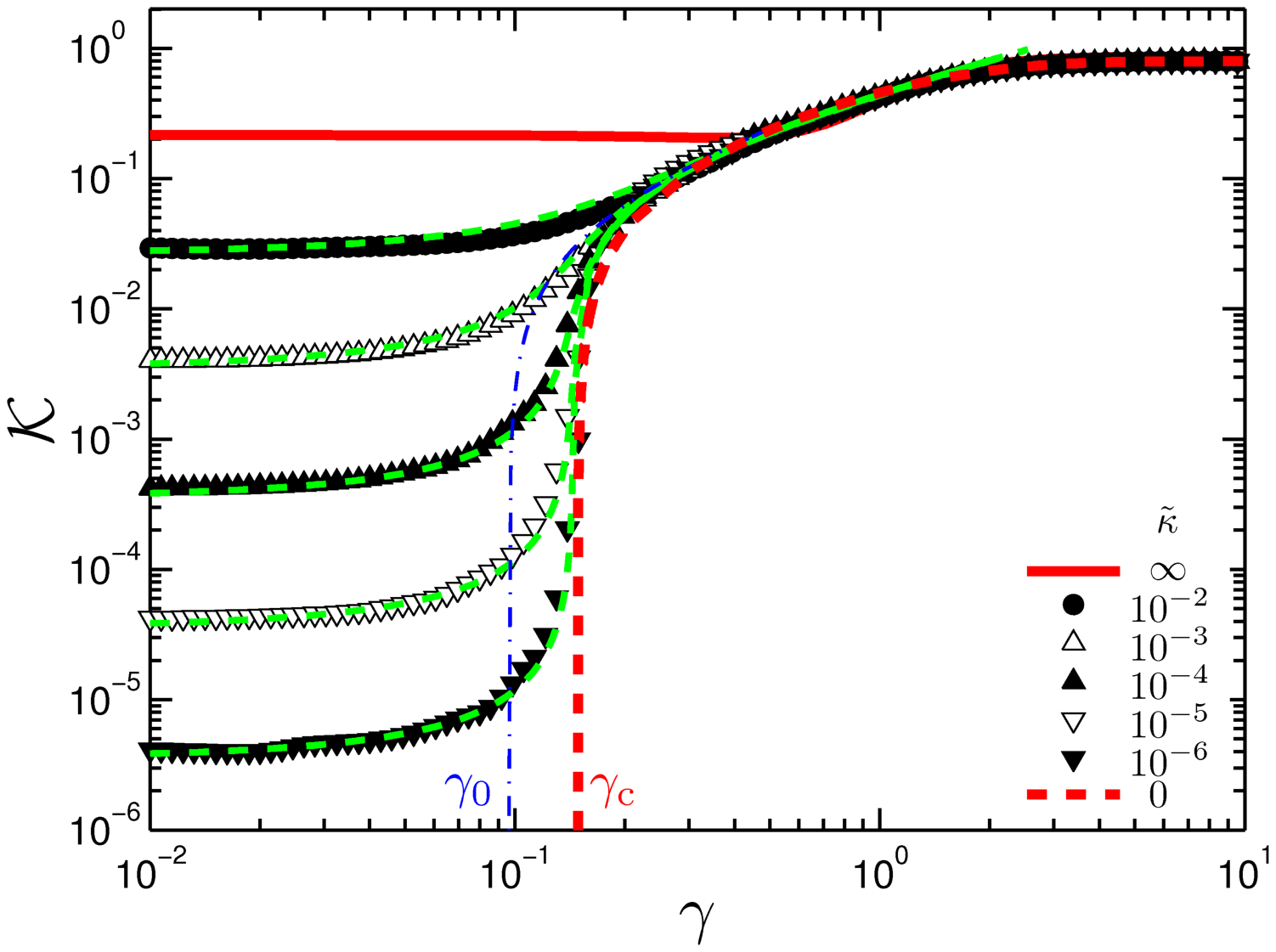} 
\includegraphics[width = \columnwidth]{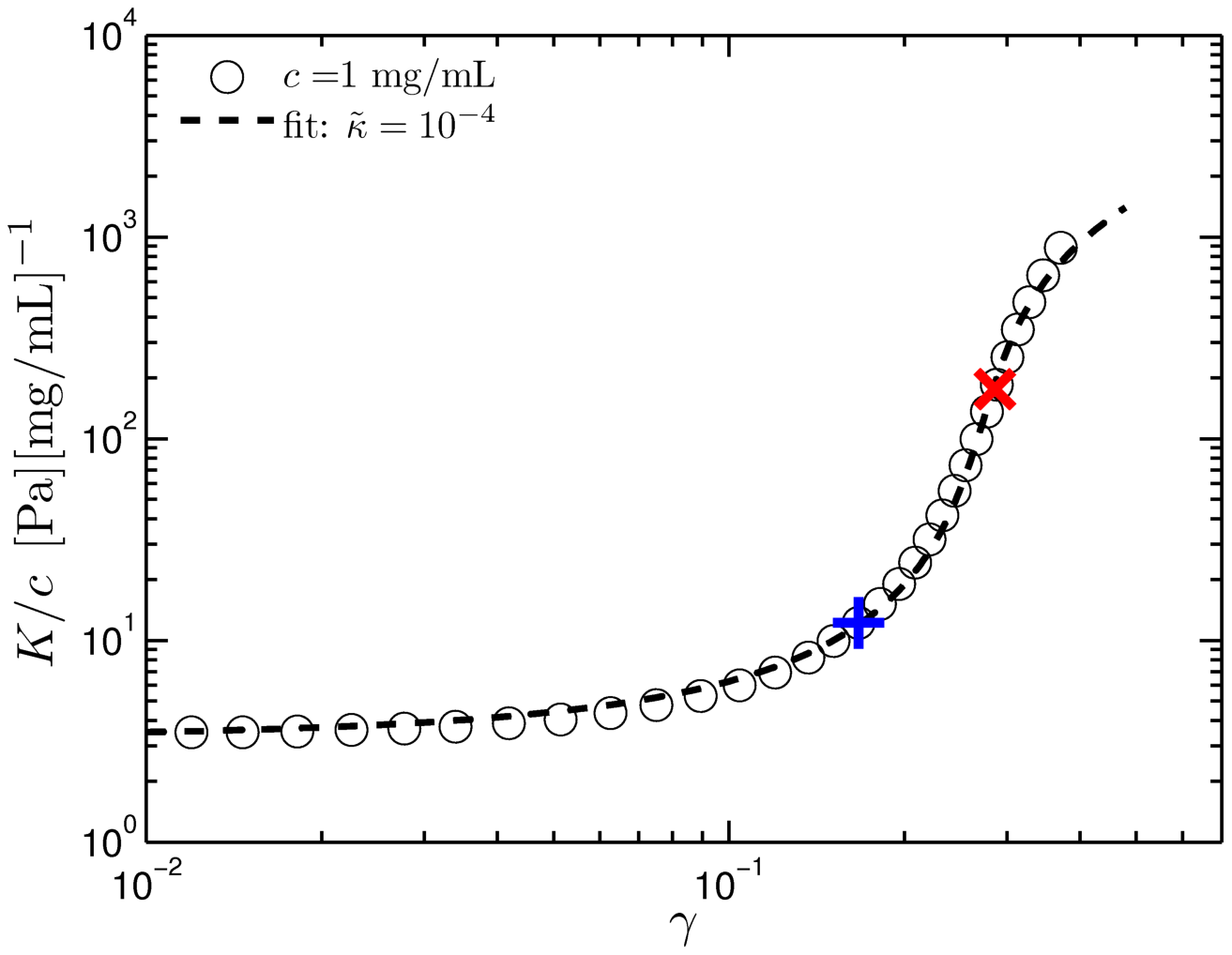}
\end{tabular}
\begin{picture}(0,0)
\put(-210,-30){\includegraphics[height=3.1cm]{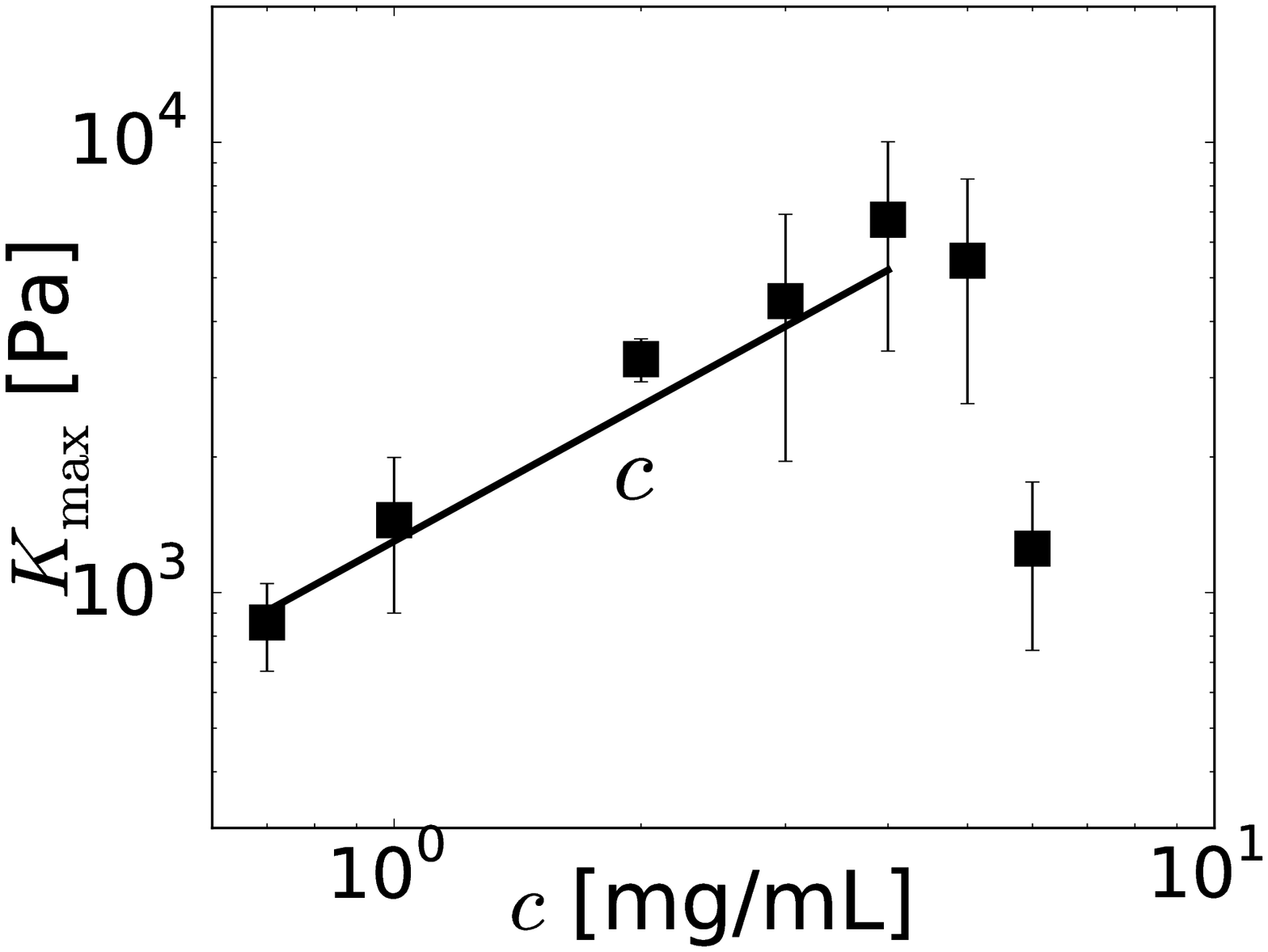}}
\end{picture}
\put (-350,-50) {\large (a)}
\put (-100,-50) {\large (b)}
\caption{(Color online) (a) Shear stiffness versus shear strain curves obtained from a phantom triangular lattice in 2D with $\langle z \rangle \simeq 3.4$. Different curves are obtained by varying the reduced bending rigidity $\tilde{\kappa}$. The onset strain for stiffening $\gamma_0$ is shown as the blue dash-dotted line. The red dashed line shows the stiffness when $\tilde{\kappa} = 0$. In absence of bending interactions, the stiffness remains zero for $\gamma \leq \gamma_c$. The green dashed lines through the symbols show the predicted stiffness according to Eq.~\eqref{crossover} with $f=0.8 \pm 0.05$ and $\phi = 2.1 \pm 0.2$. %The inset shows the scaling of linear modulus with $\tilde{\kappa}$. 
(b) Experimentally obtained stiffness versus strain curve for a 1mg/mL collagen network. Since $\mathcal{K}$ in simulations corresponds to $K/c$ in experiments, the experimentally obtained stiffness is normalized to the concentration $c$. The dashed line through the experimental data is fit according to Eq.~\eqref{crossover} with the parameters $f = 0.8$, $\phi = 2.3$ obtained from the collapse of stiffness curves obtained from simulations as explained in Sec.~\ref{criticality}. 
The critical strain $\gamma_c = 0.29$, marked with a red cross, is obtained as the inflection point of the stiffness curve. The onset strain for stiffening $\gamma_0$ is marked with a blue cross. The inset shows the experimentally measured $K_{\rm max}$ versus concentration $c$ for collagen networks, $K_{\rm max}$ is the maximum nonlinear modulus before the network ruptures. At large strains, when network stiffness is governed by stretching, the network stiffness scales as $K_{\rm max} \sim c$ shown as the black line.
}
\label{stiffeningcurves}
\end{figure*}

Here we follow up on this intriguing finding of strain-driven criticality by performing a detailed study of the nonlinear mechanics under simple shear. As a hallmark signature of criticality we demonstrate the divergence of strain fluctuations in the thermodynamic limit using finite size scaling. In Ref.~\cite{sharma2016strain} it was shown that the critical exponents appear to depend on the average connectivity in the network. Here we present our findings on the evolution of critical exponents in more detail. As another probe of criticality, we examine whether subisostatic networks exhibit critical slowing down near the critical strain. Using a simplified model of network dynamics we find evidence for power-law dynamics near the critical point.  

The article is organized as follows. In Sec.~\ref{model} we describe the computational model used in this study. We also describe the mapping of parameters used in simulations to the experimentally relevant control variables. In Sec.~\ref{criticality} we focus on the demonstration of strain-driven criticality in disordered networks. We show the critical scaling of the order parameter close to the critical point implying the continuous transition. In this section, we also analyse the stiffness versus strain curves for finite bending rigidities in terms of a crossover function. In Sec.~\ref{Divergent fluctuations}, we investigate strain fluctuations at the critical point and demonstrate their divergence in the thermodynamic limit. In Sec.~\ref{crossoverequation}, we derive an approximate equation describing the shape of the stiffness versus strain curves. We show that the derived equation can accurately describe the mechanical response measured for reconstituted collagen networks. In Sec.~\ref{deviation}, we obtain and experimentally validate scaling relation between the onset strain for stiffening and the critical strain. In Sec.~\ref{exponentsevolution}, we show that under simple shear, the critical exponents vary with the average connectivity. In Sec.~\ref{slowdynamics}, we show that the dynamics of network relaxation are critically slowed down near the critical strain for simple shear. We discuss our findings together with an outlook in Sec.~\ref{conclusions}.

\section{The Model} \label{model}
We model lattice-based networks~\cite{broedersz2011molecular, broedersz2012filament, mao2013elasticity} in 2D and 3D. Fibers are arranged on a triangular lattice (2D) or a face-centered cubic lattice (3D) of linear dimension $W$. In 2D, we randomly select two of the three fibers at each vertex on which we form a binary cross-link, i.e., enforcing local 4-fold connectivity of the network in which the third fiber does not interact with the other two~\cite{broedersz2011molecular}. Similarly, in 3D, where there are 6 fibers crossing at a point, we randomly connect three separate pairs of fibers at each vertex with binary cross-links to enforce local 4-fold connectivity~\cite{broedersz2012filament}. In both 2D and 3D, the average connectivity is further reduced below 4 by random dilution of bonds with a probability ($1-p$), where $p$ is the probability that a bond exists. The resulting connectivity after dilution can be estimated as $\langle z \rangle \simeq 4p$.
All networks, by construction, are subisostatic and floppy in the absence of bending interactions~\cite{chase2011}. The filaments are characterized by both a stretching modulus, $\mu$, and bending rigidity, $\kappa$. These define a dimensionless rigidity $\tilde\kappa = \kappa/\mu l^2$, where $l$ is the lattice spacing (mesh size) in lattice-based (Mikado) networks. In lattice-based networks we take $l=l_0$ where $l_0$ is the lattice constant. The networks are subjected to an affine simple shear strain $\gamma$ and subsequently allowed to relax by minimization of the total elastic energy.  The total elastic energy per unit volume, $\cal{H}$, is calculated using a discrete form of the extensible wormlike chain Hamiltonian~\cite{head2003distinct}
\begin{equation}\label{WLC}
\mathcal{H}=\frac{1}{W^d}\mathlarger{\mathlarger{\sum_f}} \left[\frac{\mu}{2}\int_f\left(\frac{dl}{ds}\right)^2ds+\frac{\kappa}{2}\int_f\left|\frac{d\hat{t}}{ds}\right|^2ds\right],
\end{equation}
where the term in the square brackets represents the energy stored in a single fiber and the sum is performed over all the fibres in the networks. There are other choices of modelling an individual fiber such as a truss, Euler-Bernoulli or Timoshenko beam~\cite{,huisman2007three,shahsavari2012model}. The Hamiltonian in Eq.~\eqref{WLC} captures the semiflexible nature of biopolymers with finite resistance to both tension and bending. Details about discretization of the Hamiltonian in Eq.~\eqref{WLC} are described elsewhere~\cite{licup2015elastic}. The stress and modulus are obtained by taking first and second derivatives of the energy density with respect to the applied deformation, respectively. The elastic energy involves a summation over all fibres in the network and is a function of the strain $\gamma$ and the reduced bending rigidity $\tilde{\kappa}$. Since the modulus $K$ involves the energy per unit volume, $K$ is naturally proportional to the line density $\rho$ defined as the total length of the fibers per unit volume~\cite{head2003distinct,wilhelm2003elasticity,conti2009cross,sharma2013elastic,heidemann2015elasticity}. The modulus can therefore be expressed as
\begin{equation}
K = \mu \rho \mathcal{K}\left(\gamma, \tilde{\kappa}\right),
\label{reducedmodulus}
\end{equation}
where $\mathcal{K}$ is a function of the reduced bending rigidity and the applied deformation. 
From the computational perspective, the most relevant quantity is the function $\mathcal{K}\left(\gamma, \tilde{\kappa}\right)$. Consistent with our previous studies~\cite{sharma2016strain,licup2015elastic,licup2015stress}, we report the modulus (stress) in units of $\mu \rho$. The line density $\rho$ is specific to the chosen network architecture, i.e., the network geometry. In lattice-based networks, $\rho_d=\tilde{\rho}_d/l_c^{d-1}$ with $\tilde{\rho}_\mathrm{2D}=\tfrac{6p}{\sqrt{3}}$ and $\tilde{\rho}_\mathrm{3D}=\tfrac{12p}{\sqrt{2}}$~\cite{licup2015elastic}. For Mikado networks, because of the polydispersity of $l_c$ it is more convenient to express the line density in terms of fiber length $L$ such that $\rho_\mathrm{M}=\tilde{\rho}_\mathrm{M}/L$, where $\tilde{\rho}_\mathrm{M}=n_f L^2$ and $n_f$ is the number of rods per unit area~\cite{HeadPRE2003}.

\subsection{Relationship between model and experimental parameters}\label{modelmap}
\begin{figure}[t]
\begin{tabular}{c}
\includegraphics[width = \columnwidth]{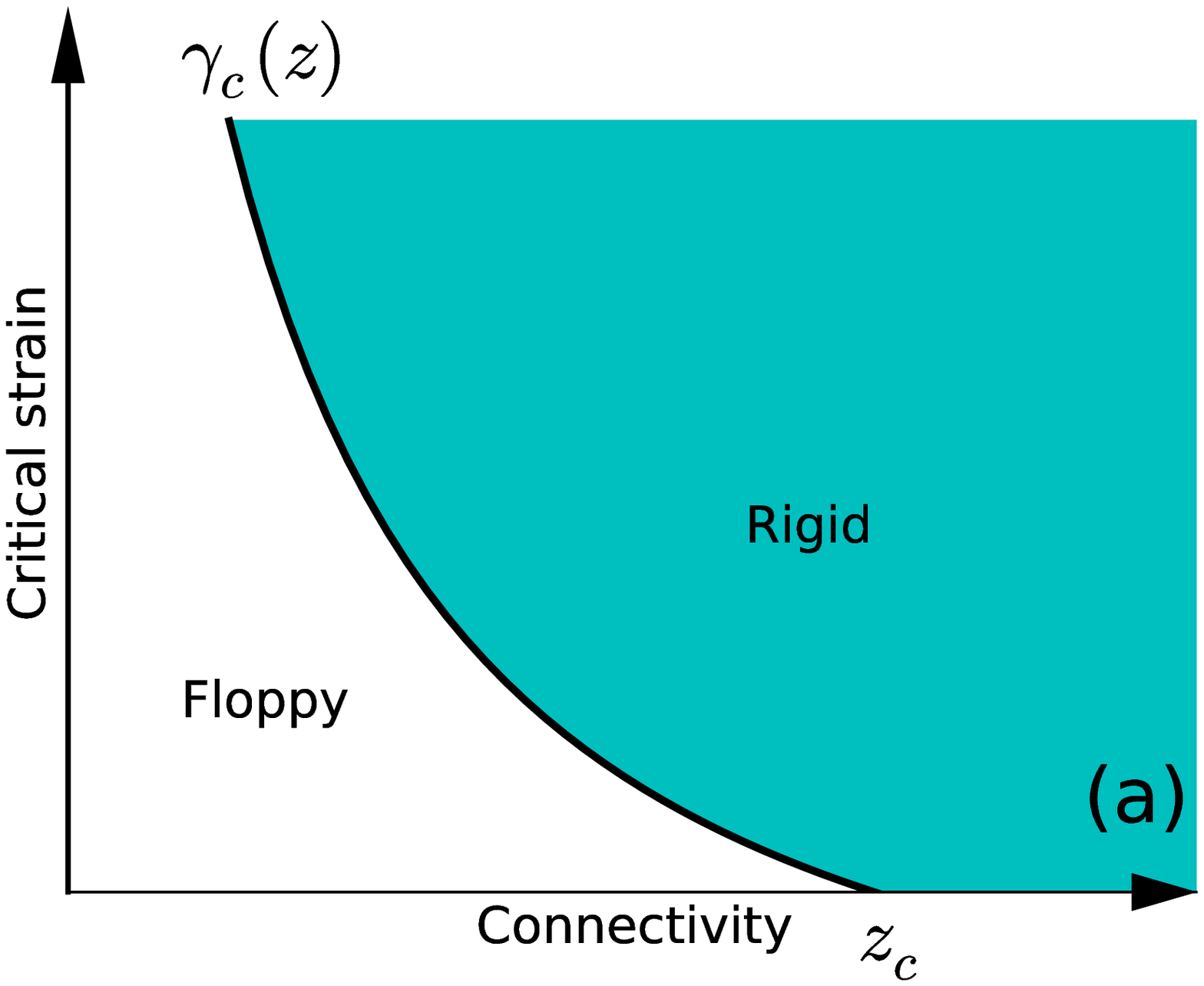} \\
\includegraphics[width = \columnwidth]{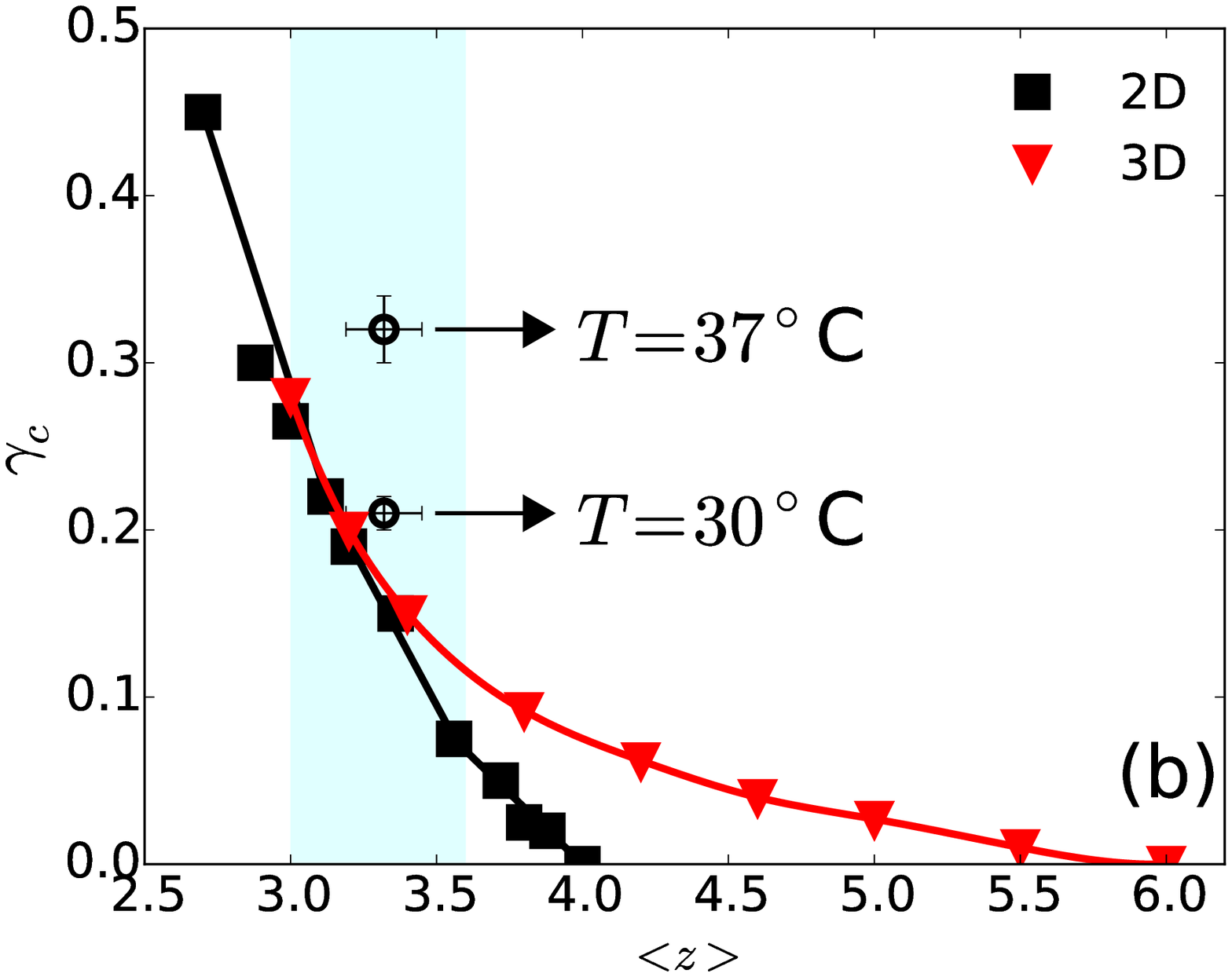}
\end{tabular}
\caption{(Color online) (a) Schematic diagram of the phase behavior of disordered fibrous networks. The curve $\gamma_c(z)$ is the boundary between floppy and rigid states. (b) $\gamma_c$ versus average connectivity for phantom triangular networks in 2D and FCC lattice based 3D networks. The critical strain decreases with increasing connectivity and approaches zero at the isostatic threshold $\langle z \rangle = 2d $. The shaded region spans connectivities in the range $3.0-3.6$. The two open symbols correspond to $\gamma_c$ of collagen networks prepared at 4mg/mL for two different polymerization temperatures. The symbols show average of 3 samples and error bars represent standard deviations. Per sample at least 100 junctions were measured to determine $\langle z \rangle$.}
\label{gcversusz}
\end{figure}

In order to map our model onto experimental parameters, we make three basic assumptions: (1) the filaments are athermal, (2) the filaments behave as rods with a homogenous elasticity, and (3) the network connectivity remains below the isostatic threshold throughout the range of polymerization conditions. Collagen networks, in general, satisfy these assumptions. Collagen fibers are rather thick and thermal fluctuations are therefore unlikely to play a significant role. As for the network connectivity, we have experimentally verified for the concentration range $0.5-4$ mg/mL and at two temperatures $T=30^{\circ}$ and $37^{\circ}$C that it remains below the isostatic threshold (see Fig.~\ref{gcversusz}(b)).

The most relevant experimental control variable is the total protein concentration $c$. For a given thickness of fibers, the volume fraction $\varphi$ of a network scales linearly with $c$ and using the above assumptions can be simply related to the reduced bending rigidity $\tilde{\kappa}$ as $\varphi \sim \tilde{\kappa}$~\cite{licup2015stress,van2016uncoupling,sharma2016strain,vahabi2016elasticity}. It follows that $K/\varphi$ (or $K/c$) in experiments can be directly compared with $\mathcal{K}\left(\gamma, \tilde{\kappa}\right)$ in simulations.

%For an elastic rod~\cite{landau1975elasticity} of radius $a$ and Young's modulus $E$, $\mu = \pi a^2 E$, $\kappa = \pi a^4 E/4$ and the line density $\rho = \frac{\varphi}{\pi a^2}$. Using these relations we obtain the normalizing factor in experiments as $\mu \rho = E \varphi$. Since $E$ is a material constant, it follows that $K/\varphi$ (or $K/c$) in experiments can be directly compared with $\mathcal{K}\left(\gamma, \tilde{\kappa}\right)$ in simulations. The parameter $\tilde{\kappa}$ in $\mathcal{K}$ is naturally related to the total protein concentration in the experiments as follows. On substituting the expressions for $\mu$ and $\kappa$, we obtain $\tilde{\kappa} \propto (a/l)^2\propto \varphi$. Hence, the volume fraction $\varphi$ (or $\rho$) in experiments can be simply related to the reduced bending rigidity $\tilde{\kappa}$ as $\rho \sim \tilde{\kappa}$.

Our theoretical results depend on the bending rigidity through the parameter $\tilde{\kappa}$ which, as shown above, scales linearly with the protein concentration in experiments. This has an important consequence for the experimental rheology results; the magnitude of modulus and stress as well as the functional dependence of the stiffness on the applied deformation are insensitive to the fibril thickness for a given concentration. This can be understood as follows. For a given total protein concentration, $\tilde{\kappa} = \kappa/\mu l^2$ is insensitive to changes in fibril thickness since $\kappa \propto a^4$, $\mu \propto a^2$ and $l \propto a$. The structure of collagen networks, including fibril thickness, mesh size, homogeneity, and presumably connectivity, depends in detail on concentration and polymerization conditions in nontrivial ways~\cite{achilli2010tailoring,hall2013toward}. However, under the basic assumptions mentioned above, $\tilde{\kappa}$ remains a constant.

\begin{figure}[t]
\includegraphics[width = \columnwidth]{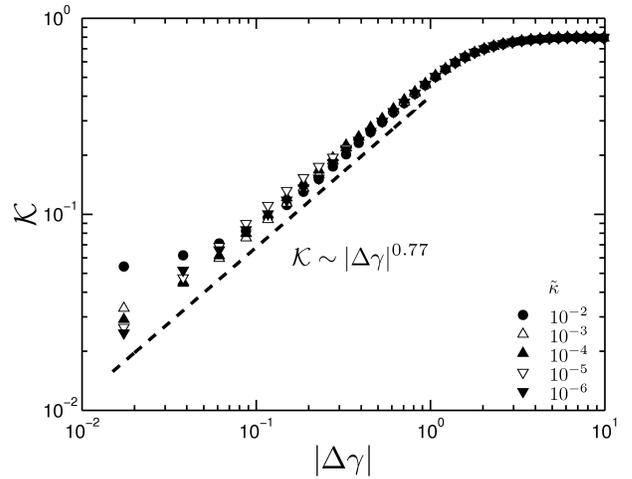}
\caption{Shear stiffness $\mathcal{K}$ versus $\Delta \gamma = \gamma - \gamma_c$ for different $\tilde{\kappa}$ obtained from simulations on a phantom triangular network in 2D with $\langle z \rangle \simeq 3.4$. In the limit of $\tilde{\kappa} \rightarrow 0$, the stiffness $K$ increases as a power-law in $\Delta \gamma$ with the critical exponent $f$, which for the given network is $\simeq 0.77$. }
\label{continuous}
\end{figure}

\section{Strain driven criticality} \label{criticality}
In Fig.~\ref{stiffeningcurves}(a), we show the network stiffness $\mathcal{K}$ as a function of the applied strain $\gamma$ for different values of $\tilde{\kappa}$. In the inset to Fig.~\ref{stiffeningcurves}(a), we show that the linear modulus scales linearly with $\tilde{\kappa}$. The scaling $\mathcal{K} \sim \tilde{\kappa}$ in the linear regime has been reported in several computational studies~\cite{Head2003PRL,wilhelm2003elasticity,wyart2008elasticity,chase2011,picu2011mechanics, licup2015stress,sharma2016strain}. That the computational model is suitable for studying athermal networks such as collagen is based on the following observations. (1) The computationally obtained modulus is in units of $\mu \rho$ (Eq.~\eqref{reducedmodulus}) implying that $G_0 \equiv K(\gamma=0) \sim \rho \tilde{\kappa} \sim \rho^2$ consistent with experimental data sets on reconstituted networks of collagen type I~\cite{sharma2016strain, licup2015stress, motte2013strain,piechocka2011rheology}. %In the inset to Fig.~\ref{stiffeningcurves}(b) we show that the experimentally obtained linear modulus of reconstituted collagen network exhibits quadratic scaling with concentration. The slight deviation from the quadratic scaling is explained later in the article. 
(2) As can be seen in Fig.~\ref{stiffeningcurves}(a) the onset of nonlinearity occurs at a strain $\gamma_0$ which appears to be independent of $\tilde{\kappa}$. Experimentally this corresponds to $\gamma_0$ being independent of the total protein concentration which is indeed what has been observed in several studies~\cite{motte2013strain,piechocka2011rheology,licup2015stress, sharma2016strain}. (3) For large strains, $\mathcal{K}$ is independent of $\tilde{\kappa}$, implying that $K \sim \rho$, which is expected in the regime where strains are large enough to cause stretching of fibers. In order to verify if experiments indeed show the linear scaling of $K$ with concentration we consider how $K_{\rm max}$ varies with the concentration $c$.  $K_{\rm max}$ is the nonlinear modulus of a network before undergoing failure due to the applied stress. As can be seen in the inset of Fig.~\ref{stiffeningcurves}(b), the experimentally measured $K_{\rm max}$ scales linearly with the concentration implying that for large strains the nonlinear stiffness scales as $K \sim \rho$.
%is consistent with the $c^{-1}$ scaling implying that at large strains $K \sim \rho$. The slight deviation from $c^{-1}$ scaling is due to the fact that $G_0/c^2$ has a weak $c$ dependence as we explained in Ref.~\cite{sharma2016strain}.
 
  % Experimentally obtained $K/c$ is shown in Fig.~\ref{stiffeningcurves}b as a function of the applied strain for different concentrations $c$. There are certain qualitative features of rheology in experiments and simulations which yield an insight into the mechanics of fiber networks: (1) In the linear regime $\mathcal{K} \sim \tilde{\kappa}$ in simulations and $K/c \sim c$ in experiments, and (2) Onset of nonlinearity occurs at a strain $\gamma_0$ which appears to be independent of $\tilde{\kappa}$ in simulations and $c$ in experiments, and lastly (3) For large strains, $\mathcal{K}$ is independent of $\tilde{\kappa}$ in simulations and $K/c$ is independent of $c$ in experiments. The scaling $\mathcal{K} \sim \tilde{\kappa}$ in the linear regime has been reported in several computational studies~\cite{Head2003PRL,wilhelm2003elasticity,wyart2008elasticity,chase2011,licup2015stress,sharma2016strain}. Since $\tilde{\kappa} \sim \rho $ as derived in Sec.~\ref{modelmap} and $\mathcal{K} \sim \rho \tilde{\kappa}$ in the linear regime, theoretically one expects the linear modulus $K \sim \rho^2$ which is in fact consistent with the experimental studies on reconstituted collagen networks~\cite{motte2013strain,piechocka2011rheology}. 

When fiber bending costs no energy, i.e., $\tilde{\kappa} = 0$, the stiffness $\mathcal{K}$ remains zero for strains $|\gamma| \leq \gamma_c$. Above $\gamma_c$, $\mathcal{K}$ increases continuously from zero for $\tilde{\kappa} = 0$. The critical strain $\gamma_c$ is determined by the network architecture, in particular its average connectivity~\cite{sheinman2012nonlinear}. At the isostatic threshold of $\langle z \rangle = 2d$, a central force network is marginally stable with $\gamma_c = 0$. In Fig.~\ref{gcversusz}(a), we show a schematic of the phase diagram in $\gamma$-$z$ plane.  For a given average connectivity below the isostatic threshold, increasing the deformation beyond $\gamma_c$ causes a phase transition from floppy to rigid phase. The continuous curve $\gamma_c(z)$ marks the boundary between the floppy and rigid states of subisostatic networks. In Fig.~\ref{gcversusz}(b) we show the computationally obtained $\gamma_c$ versus the average connectivity in the network when subjected to simple shear deformation. The critical strains for both 2D and 3D networks are in quantitative agreement as long as the connectivity is suffciently below the 2D isostatic point $\langle z \rangle = 4$. The shaded region in Fig.~\ref{gcversusz}(b) spans the connectivities relevant for collagen networks. Also shown are two values of $\gamma_c$ of 4mg/mL collagen networks for two temperatures. The critical strain is in quantitative agreement with the model for $T=30^{\circ}$C. The apparent disagreement for $T=37^{\circ}$C is probably due to the uncertainty associated with determination of $\langle z \rangle$ in the experiments. It is possible that due to finite resolution in experiments, the connectivity at some of the nodes is measured as $4$ due to overlapping collagen fibers. This would lead to an overestimation of $\langle z \rangle$ and can thus account for the disagreement  between theory and experiments at $T=37^{\circ}$C. 

%change in the network architecture with temperature~\cite{licup2015stress}. 

In Fig.~\ref{continuous}, the network stiffness is shown for several values of $\tilde{\kappa}$ in the vicinity of the critical strain. The continuous nature of the transition from floppy to rigid states is evident in the critical scaling of the network stiffness $\mathcal{K} \sim|\Delta\gamma|^f$ where $\Delta\gamma = \gamma - \gamma_c \geq 0$ and $f$ is a critical exponent. As shown in Fig.~\ref{continuous}, the power-law scaling of stiffness is apparent only in the limit of $\tilde{\kappa} \rightarrow 0$. Extracting $f$ as the limiting slope of $\mathcal{K}$ vs. $\Delta\gamma$ provides an independent method of obtaining this critical exponent. The same exponent can be obtained by scaling analysis as described in Sec.~\ref{crossover_theory}. Power-law scaling of the order parameter, $\mathcal{K}$ (or $K$) in our case, is a hallmark signature of critical phenomena. In fact, the strain-driven phase transition is strictly defined only for $\tilde{\kappa} = 0$ at which the interactions within the network are purely central force interactions. Upon addition of a field such as fiber bending, the network becomes stable for $\gamma < \gamma_c$ with the stiffness $\mathcal{K} \propto \tilde{\kappa}$.

In absence of bending interactions, the phase behavior characterized by the continuous transition of the order parameter $\mathcal{K}$ is reminiscent of the ferromagnetic phase transition. Magnetic materials are characterized by a Curie temperature $T_c$ such that for $T$ above $T_c$, the material is paramagnetic. On lowering the temperature $T$ below $T_c$, there is spontaneous magnetization $M$ of the material which increases continuously from zero as $M \propto |\Delta T|^{\beta}$ where $\Delta T = T - T_c < 0$ and $\beta$ is the critical exponent. Above the Curie temperature, the paramagnetic phase is characterized by a zero magnetization. However, in presence of a finite magnetic field $H$, there is a net magnetization in the paramagnetic phase with $M \propto H$. It is an intriguing analogy that by mapping $\tilde{\kappa}$ to external field $H$ and $\gamma$ to the temperature $T$, one can study the transition from floppy to rigid states the same way as in a ferromagnet as further elaborated in Sec.~\ref{crossoverequation}.

\begin{figure}[t]
\includegraphics[width = \columnwidth]{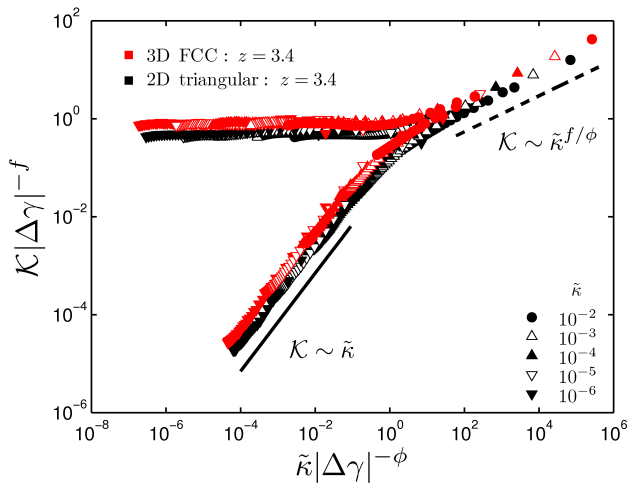}
\caption{(Color online) (a) Collapse of shear stiffness versus shear strain curves of Fig.~\ref{stiffeningcurves}(a) according to Eq.~\eqref{collapse}. Simulation data from 3D network with same connectivity as in 2D of $\langle z \rangle \simeq 3.4$ collapse with the same critical exponents $f = 0.8$ and $\phi = 2.1$.}
\label{widomcollapse}
\end{figure}

\subsection{Crossover for finite $\tilde{\kappa}$} \label{crossover_theory}
The power law scaling of $\mathcal{K}$ with $\Delta \gamma$ is a hallmark signature of criticality and is strictly observed only when $\tilde{\kappa} = 0$. It is obvious that in this regime, the modulus is entirely governed by stretching of fibers. For any finite $\tilde{\kappa}$, a subisostatic network is stable for $\Delta \gamma < 0$. In fact, for sufficiently small $\tilde{\kappa}$, the linear modulus of a subisostatic network is bending governed leading to $\mathcal{K}\sim\tilde\kappa$ for $\gamma<\gamma_c$~\cite{Head2003PRL,wilhelm2003elasticity,wyart2008elasticity,chase2011,licup2015stress,sharma2016strain}. Analogous to ferromagnetism, in presence of finite auxiliary field $\tilde{\kappa}$, the network undergoes a strain driven crossover from the bend dominated regime $\Delta \gamma < 0$  to the stretch dominated regime $\Delta \gamma > 0$. These two regimes can be summarized by the scaling form
\begin{equation}
\mathcal{K} \propto \left|\Delta \gamma\right|^f\mathcal{G_{\pm}}\left(\frac{{\tilde\kappa}}{{|\Delta \gamma|^{\phi}}}\right),
\label{collapse}
\end{equation}
where $\mathcal{G_{\pm}}$ is a scaling function with the positive and negative branches corresponding to $\Delta \gamma >0$ and $\Delta \gamma <0$, respectively. This scaling is analogous to that for the conductivity of random resistor networks and fiber networks as a function of connectivity \cite{straley1976critical,chase2011}. In Fig.~\ref{widomcollapse}, we test this by plotting $\mathcal{K}|\Delta\gamma|^{-f}$ vs. $\tilde\kappa|\Delta\gamma|^{-\phi}$, according to Eq.~\eqref{collapse}. For $x\ll 1$, $\mathcal{G_+}(x)$ is approximately constant and $\mathcal{G_-}(x)\propto x$. That $\mathcal{G_+}(x)$ is approximately constant for $x\ll 1$ captures the critical scaling of $K$ as $\mathcal{K} \sim |\Delta \gamma|^f$. The scaling $\mathcal{G_-}(x)\propto x$ captures the bend-dominated linear modulus where the linear modulus scales as $\mathcal{K} \sim \tilde{\kappa}$. Since $K$ must be finite at $\Delta \gamma = 0$, we also expect $\mathcal{K}\sim \kappa^{f/\phi}\mu^{1-f/\phi}$, consistent with Eq.~\eqref{collapse}. We show in Fig.~\ref{widomcollapse} the data obtained from phantom triangular networks in 2D (same as in Fig.~\ref{stiffeningcurves}(a)) and FCC-based 3D lattices collapsed according to Eq.~\eqref{collapse}. Interestingly, the data collapse with the same exponents $f\simeq0.8$ and $\phi\simeq2.1$. The average connectivity for the two different networks is chosen to be $\simeq 3.4$. Data from Mikado networks with the same average connectivity $\langle z \rangle$ as in lattice-based networks can be collapsed with the same critical exponents~\cite{sharma2016strain}. In fact, as we show in Sec.~\ref{exponentsevolution}, the exponents appear to be independent of the spatial dimensionality and are primarily determined by the average connectivity. 

Mapping protein concentration to $\tilde{\kappa}$ as described in Sec.~\ref{modelmap} allows us to obtain an analogous scaling relation applicable to experimental data. Since computationally one obtains $\mathcal{K}$ one must create the analogous quantity in experiments by scaling the measured modulus with concentration, i.e., $K/c$. On substituting $c$ for $\tilde{\kappa}$ and $K/c$ for $\mathcal{K}$ in Eq.~\eqref{collapse}, we obtain the scaling function to collapse the experimental data as shown by us in Ref.~\cite{sharma2016strain}.

%\begin{equation}
%\frac{K}{c}\propto \left|\Delta \gamma\right|^f\mathcal{G_{\pm}}\left(\frac{{c}}{{|\Delta \gamma|^{\phi}}}\right).
%\label{exptcollapse}
%\end{equation}
%It has been shown in Ref.~\cite{sharma2016strain}, that Eq.~\eqref{exptcollapse} can collapse experimentally obtained stiffening curves with critical exponents corresponding to an average connectivity $\simeq 3.2$.

The scaling function $\mathcal{G_{\pm}}$, with $f$ and $\phi$ as input parameters, describes the stiffening curves over the entire elastic regime for any concentration (or $\tilde{\kappa}$ in simulations). One can obtain an analytical $\mathcal{G_{\pm}}$ (approximately) exploiting the analogy of nonlinear mechanics to ferromagnetism as we show in Sec.~\ref{crossoverequation}.

\begin{figure*}
\begin{tabular}{@{}c@{}}
\includegraphics[width = \textwidth]{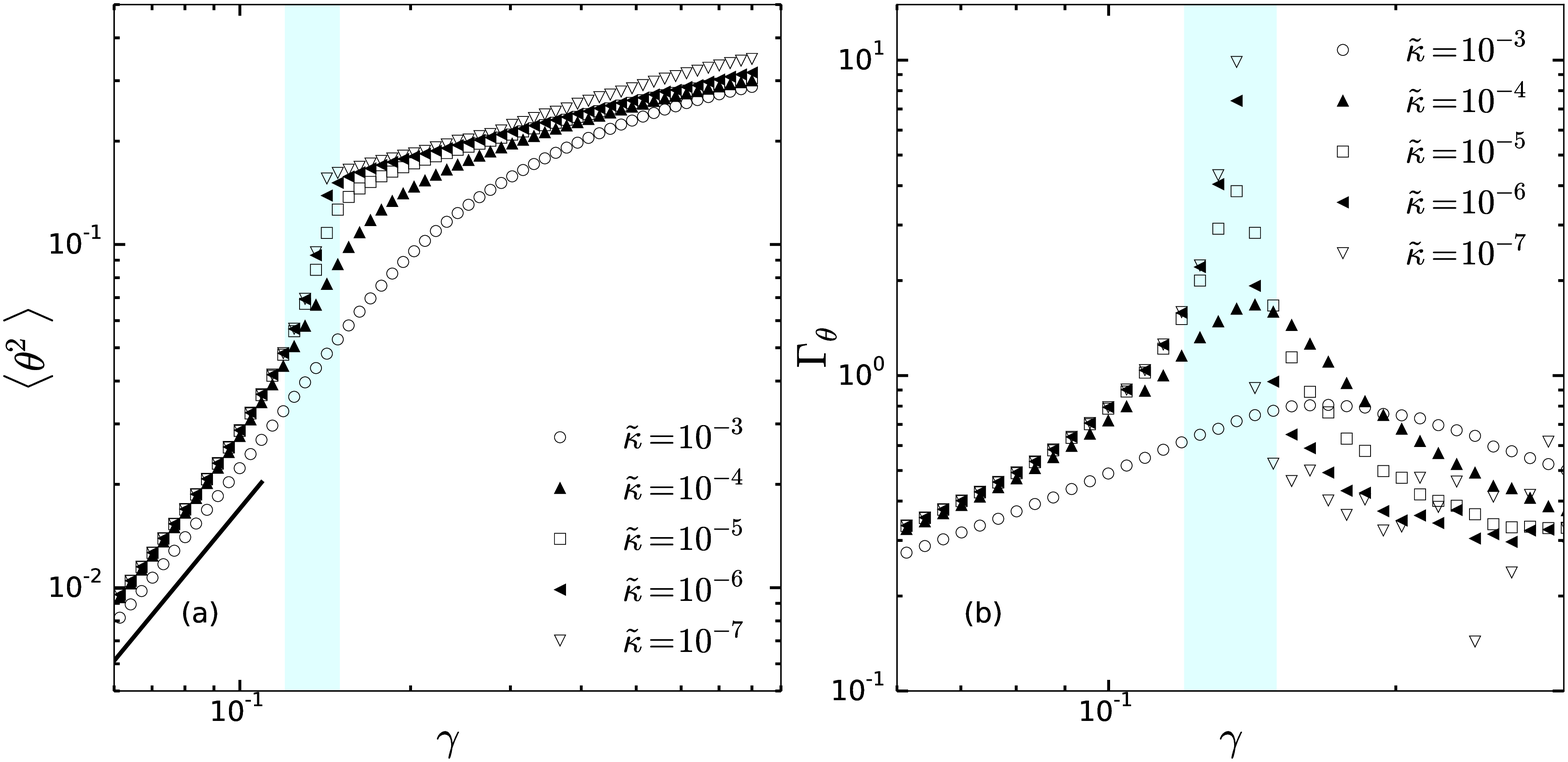} \\
\includegraphics[width = \textwidth]{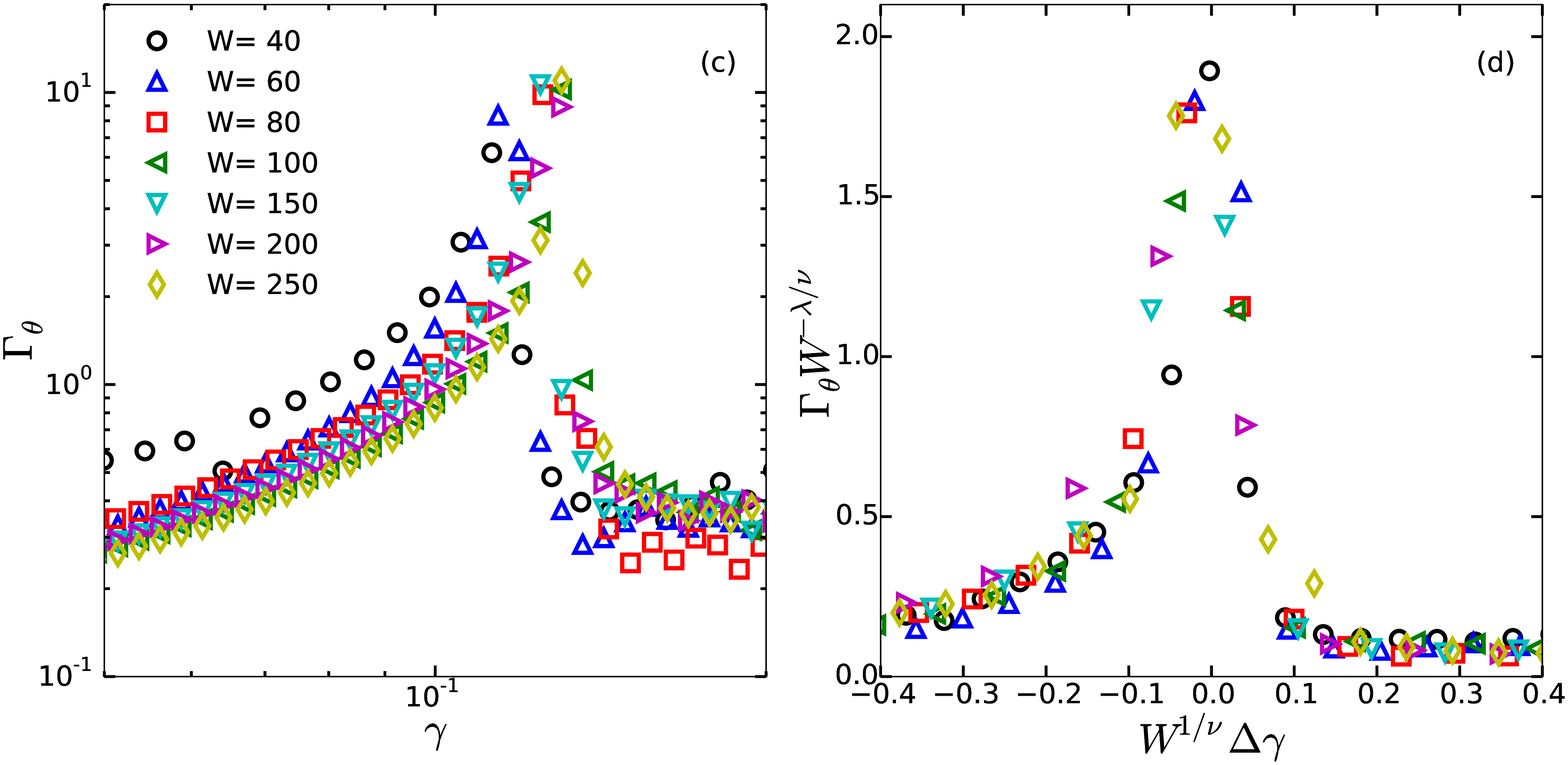} \\
\end{tabular}
\caption{(Color online) Divergent fluctuations at the critical strain. (a) Average bending angle $\langle \theta^2 \rangle$ obtained from simulations on a phantom triangular network in 2D with $\langle z \rangle \simeq 3.4$ for different values of $\tilde{\kappa}$ (see legend). The network size is $W^2 = 250^2$. The thick black line indicates the expected small-strain $\gamma^2$ scaling. $\langle \theta^2 \rangle$ increases monotonically with $\gamma$. The shaded region is approximately the range $\gamma_c - \gamma_0$. In this range, the rate of increase of $\langle \theta^2 \rangle$ is strongly dependent on $\tilde{\kappa}$. (b) $\Gamma_{\theta}(\gamma)$ obtained as the derivative of data in (a) with respect to $\gamma$. In the limit of $\tilde{\kappa} \rightarrow 0$, $\Gamma_{\theta}$ diverges at $\gamma = \gamma_c$. (c) $\Gamma_{\theta}$ versus $\gamma$ for different system sizes (see legend). The bending rigidity is $\tilde{\kappa} = 10^{-7}$. (d) Collapse of data in (c) according to Eq.~\eqref{fss} with $\lambda = 0.6 \pm 0.1$ and $\nu = 2.0 \pm 0.1$.
}

%\caption{(Color online) Divergent fluctuations at the critical strain. (a) $\Gamma_{\theta}$ obtained from simulations on a phantom triangular network in 2D with $\langle z \rangle \simeq 3.4$, as the inflection point of the $\log K$ vs.\ $\log \gamma$ curves. The curves correspond to different values of $\tilde{\kappa}$ (see legend). In the limit of $\tilde{\kappa} \rightarrow 0$, the rate of increase of bending in the network, close to the critical strain diverges. (b) Differential non-affinity $\delta \Gamma(\gamma)$ obtained from the same simulations in (a) as a function of the applied strain. $\delta \Gamma(\gamma)$ peaks at $\gamma \simeq \gamma_c$. As expected, the height of the peak increases with decreasing $\tilde{\kappa}$ since the displacement field of these networks becomes highly non-affine as $\gamma\rightarrow\gamma_c$. Non-affine displacements  in a network with $\tilde{\kappa} = 10^{-6}$ are shown in the lower panel as the network is deformed through the critical strain $\gamma_c$. The arrows indicate the deviation of a node from the imposed deformation. The magnitude of the vectorial displacements is largest at the critical strain. The color bar on the right indicates the elastic energy in bending (green) or stretching (red) form.}
\label{divergentfluc}
\end{figure*}

\begin{figure*}[t]
\includegraphics[width = \textwidth]{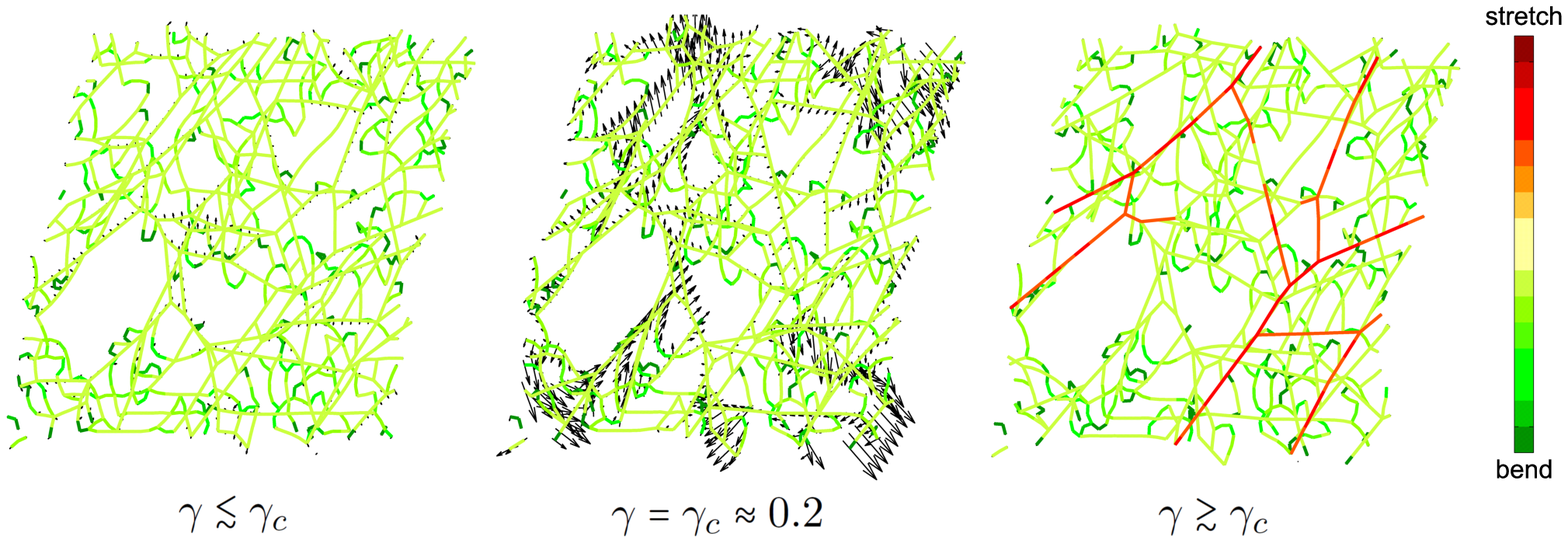}
\caption{(Color online) Non-affine displacements in a 2D phantom triangular network with $\langle z \rangle \simeq 3.4$ and $\tilde{\kappa} = 10^{-6}$ are shown as the network is deformed through the critical strain $\gamma_c$. The arrows indicate the deviation of a node from the imposed deformation. The magnitude of the vectorial displacements is largest at the critical strain. The color bar on the right indicates the elastic energy in bending (green) or stretching (red) form.}
\label{diffNA}
\end{figure*}

\subsection{Divergent fluctuations}\label{Divergent fluctuations}
In a thermal critical phenomenon, there are divergent fluctuations in the order parameter at the critical point. In the athermal network under consideration in this study, there are no divergent fluctuations in the macroscopic $\mathcal{K}$.  One can, however, measure fluctuations by considering the deviation of the strain field within the network from the expected affine field~\cite{hatami2008scaling,heidemann2015elasticity}. Under affine deformation, filaments are either stretched or compressed. Deviations from the affine deformation induce bending on filaments which can be considered as a measure of fluctuations. These fluctuations are suppressed by a finite field such as $\tilde{\kappa}$. In Fig.~\ref{divergentfluc}(a), we plot the bending angle $\theta_{ijk}$ averaged over the entire network for different values of $\tilde{\kappa}$. The triplet $\{i,j,k\}$ corresponds to three consecutive crosslinks labeled as $i$, $j$, and $k$ and the average implies summing over all the triplets in the network. As can be seen in Fig.~\ref{divergentfluc}(a), the average bending angle increases with the applied deformation. For small strains, the increase is quadratic in $\gamma$ as expected in the linear regime. At large strains, the average bending angle increases very slowly with the applied deformation. In the intermediate strain range, shown as the shaded region, the rate of increase of average bending angle depends strongly on $\tilde{\kappa}$. We define $\Gamma_{\theta}$ as the rate of change  of the average bending angle with the applied strain.
\begin{equation}
\Gamma_{\theta}(\gamma) = \frac{\partial \langle \theta_{ijk}^2\rangle}{\partial \gamma}.
\label{gtheta}
\end{equation}  
In Fig.~\ref{divergentfluc}(b), we plot $\Gamma_{\theta}$ as a function of $\gamma$ for different values of $\tilde{\kappa}$. These results are obtained from simulations on a phantom triangular network in 2D with $\langle z \rangle \simeq 3.4$. The maximum of  $\Gamma_{\theta}$ shifts to the left in $\gamma$ with decreasing bending rigidity. In the limit of $\tilde{\kappa} = 0$, the peak height is maximum for a given network size and it is located at the critical strain $\gamma = \gamma_c$.

The quantity $\Gamma_{\theta}$ is expected to diverge in the thermodynamic limit at $\gamma  = \gamma_c$ for $\tilde{\kappa} = 0$. In Fig.~\ref{divergentfluc}(c), we show $\Gamma_{\theta}$ for different system sizes $W$. These curves are obtained for a fixed small $\tilde{\kappa} = 10^{-7}$. If $\Gamma_{\theta}$ diverges as $|\gamma - \gamma_c|^{-\lambda}$ in the thermodynamic limit $W\rightarrow \infty$, then the following scaling relation must capture the scaling behavior of $\Gamma_{\theta}$ for finite $W$:
\begin{equation}
\Gamma_{\theta} \propto W^{\lambda/\nu}\mathcal{H}(W^{1/\nu} \Delta \gamma),
\label{fss}
\end{equation}
where $\nu$ is the exponent associated with the divergence of correlation length~\cite{sharma2016strain}, $\Delta \gamma = \gamma - \gamma_c$ is the distance from the critical strain and $\mathcal{H}(x)$ is a scaling function. We show in Fig.~\ref{divergentfluc}(d), the collapse of data in Fig.~\ref{divergentfluc}(c) according to Eq.~\eqref{fss} with the exponents $\lambda = 0.6 \pm 0.1$ and $\nu = 2.0 \pm 0.1$. With these exponents, the peak height of $\Gamma_{\theta}$ is expected to scale as $W^{\lambda/\nu} \sim W^{0.3}$. It follows that due to the weak system size dependence, a clear demonstration of $W^{\lambda/\nu}$ scaling of the peak height requires much larger system sizes than those studied in this work. Nevertheless, the collapse in Fig.~\ref{divergentfluc}(d) provides convincing evidence for $\Gamma_{\theta}$ as an appropriate measure of fluctuations in fibrous networks.

Another measure of fluctuations is the differential non-affinity which measures the strain fluctuations within the network. Given the displacement field $\mathrm{\mathbf{u}}$ and the \emph{affine} displacement field $\mathrm{\mathbf{u}^{A}}$ of the network, the non-affine fluctuations can be quantified as~\cite{sheinman2012nonlinear}
\begin{equation}
\delta \Gamma(\gamma) = \frac{\langle \|\delta \mathrm{\mathbf{u}^{NA}}\|^2 \rangle}{l^2 d\gamma^2},
\label{dna}
\end{equation}
where $\delta \Gamma(\gamma)$ is referred to as differential non-affinity, $\delta\mathrm{\mathbf{u}^{NA} = \mathbf{u} - \mathbf{u}^{A}}$ is the differential non-affine displacement of a crosslink to an imposed strain $d\gamma$, $l$ is the typical network mesh size and the angular brackets represent a network average. In Ref.~\cite{sharma2016strain}, we showed that $\delta \Gamma(\gamma)$ exhibits a peak at $\gamma = \gamma_c$, the height of which increases with decreasing $\tilde{\kappa}$. In Fig.~\ref{diffNA}, we show the differential non-affine displacements $\delta\mathrm{\mathbf{u}^{NA}}$ superimposed on network nodes in the neighborhood of $\gamma_c$. The magnitude of non-affine displacements is largest at the critical strain. It follows that the network is at its most susceptible mechanical state at $\gamma=\gamma_c$ requiring large scale internal rearrangements in response to an infinitesmal external deformation. The nature of deformation within the network changes dramatically when the applied deformation increases through $\gamma = \gamma_c$. Whereas the network deforms primarily through bending modes for $\gamma \leq \gamma_c$, stretching becomes the dominant deformation mode for $\gamma > \gamma_c$.

%We calculate both measures of fluctuations (Eqs.~\eqref{gtheta} and \eqref{dna}) in the network for different $\tilde{\kappa}$ and study the behavior in the limit of $\tilde{\kappa} \rightarrow 0$. For finite $\tilde{\kappa}$, we define the critical strain $\gamma_c$ as the inflection point of strain-stiffening curves plotted as the $\log \mathcal{K}$ vs.\ $\log \gamma$ curves, analogous to the determination of the critical point in a finite-size system~\cite{stauffer1994introduction}. In Fig.~\ref{divergentfluc}(a), we show $\Gamma_{\theta}$ versus $\gamma$ obtained from simulations on a phantom triangular network in 2D with $\langle z \rangle \simeq 3.4$, for different values of the fibre bending rigidity. The maximum of  $\Gamma_{\theta}$ shifts to the left in $\gamma$ with decreasing bending rigidity. In the limit of $\tilde{\kappa} = 0$, the position of the maximum coincides with the critical strain $\gamma_c$. It can be clearly seen in Fig.~\ref{divergentfluc}(a) that the bending angles increase steeply as $\gamma \rightarrow \gamma_c$. In Fig.~\ref{divergentfluc}(b), we show $\delta \Gamma(\gamma)$ in the neighbourhood of the critical strain. The fluctuations grow with decreasing $\tilde{\kappa}$, consistent with the idea that the bending rigidity can be considered as an auxiliary field. 

Finite-size scaling analysis of the order parameter $\mathcal{K}$ reveals underlying divergence of the correlation length as shown in Ref.~\cite{sharma2016strain}. The diverging correlation length, together with divergent fluctuations and the continuously evolving order parameter constitute evidence in favor of a second-order type strain-driven phase transition in disordered networks.

\section{Equation for the crossover function} \label{crossoverequation}
The scaling ansatz and function $\mathcal{G}_{\pm}(x)$ in Eq.~\eqref{collapse} can account well for the nonlinear mechanics of our model networks for any $\tilde{\kappa}$ and $\gamma$. We can obtain an analytical approximation for $\mathcal{G}_{\pm}(x)$ in a way analogous to the approach for ferromagnetism~~\cite{ArrottPRL1967,goldenfeld1992lectures}. In a way similar to the equation of state relating magnetic field $H$ to magnetization $M$, we postulate the following mean-field equation of state for bending stiffness $\tilde\kappa$ and as a series in the shear modulus $\mathcal{K}$~\cite{broedersz2014modeling}:
\begin{equation}
\tilde\kappa \sim b\mathcal{K}+c\mathcal{K}^2,
\end{equation}
where $b\sim\Delta\gamma$ for a transition controlled by strain. Here, in contrast with the order parameter $M$ for ferromagnetism, symmetry does not forbid a quadratic term in this equation of state~\cite{broedersz2014modeling}. After a minor change in normalization, this can be rewritten as 
\begin{equation}
\frac{\tilde\kappa}{|\Delta\gamma |^2} \sim \frac{\mathcal{K}}{|\Delta\gamma|}\left(\mp1+\frac{\mathcal{K}}{|\Delta\gamma|}\right),
\end{equation}
where the upper `$-$' refers to $\gamma>\gamma_c$ and the lower `$+$' refers to $\gamma<\gamma_c$.
This yields $\mathcal{K}\sim |\Delta\gamma|$ for small $\Delta\gamma>0$ and $\tilde\kappa=0$, while $\mathcal{K}\sim\tilde\kappa$ for $\Delta\gamma<0$ and small $\tilde\kappa>0$. As shown above, our results deviate from the mean-field behavior, $\mathcal{K}\sim |\Delta\gamma|^f$, where $f=1$. We find $f\simeq0.8$.

As is done for ferromagnetism, the equation of state above can be written in a form that can account for non-mean-field exponents, while remaining non-singular except at the critical point ($\Delta\gamma=\tilde\kappa=0$). We introduce potentially non-integer exponents $f$ and $\phi$, where
\begin{equation}
\frac{\tilde{\kappa}}{|\Delta \gamma|^{\phi}} \sim \frac{{\mathcal{K}}}{|\Delta \gamma|^f}\left(\mp 1+\frac{{\mathcal{K}}^{1/f}}{|\Delta \gamma|}\right)^{(\phi - f)}.
\label{crossover}
\end{equation}
For $\Delta \gamma = 0$, this scaling relation corresponds to $\mathcal{K} \sim \tilde{\kappa}^{f/\phi}$ at the critical point. Again, the mean-field values of the exponents are $f=1$ and $\phi=2$.

Equation~\eqref{crossover} can be used to calculate $\mathcal{K}$ for any $\gamma$. The input parameters are $\tilde{\kappa}$, $f$, $\phi$ and $\gamma_c$.  The critical strain $\gamma_c$ can be independently determined from a network with only central-force interactions. The critical exponents are obtained from the data collapse using Eq.~\eqref{collapse}. In Fig.~\ref{stiffeningcurves}(a), we use Eq.~\eqref{crossover} to obtain $\mathcal{K}$ as a function of $\gamma$ for different $\tilde{\kappa}$. The stiffening curves calculated using Eq~\eqref{crossover} are shown together with the numerically obtained curves. Clearly, Eq.~\eqref{crossover} can accurately predict the nonlinear stiffening curves. 

Equation~\eqref{crossover} can accurately capture the experimentally obtained stiffening curves of collagen networks~\cite{sharma2016strain}. However, the fitting procedure, when applied to experiments needs to be slightly modified. The fitting to experimental data is done in the following way. We first focus on the linear regime. In the linear regime, we know from simulations that the modulus (in units of $\rho \mu$) scales linearly with $\tilde{\kappa}$ which itself scales as $\tilde{\kappa} \sim \rho$ giving rise to a $c^2$ (or $\rho^2$) dependence of the linear modulus where $c$ is the protein concentration. However, as shown in the inset of Fig.~\ref{stiffeningcurves}(b), the linear modulus obtained experimentally from reconstituted collagen networks exhibits $K \sim c^{2+\delta}$ scaling. It is plausible that the deviation from the $c^2$ scaling is simply a consequence of experimental uncertainties. However, as shown in Ref.~\cite{sharma2016strain}, the deviation from $c^2$ scaling is probably due to the weak dependence of $\gamma_c$ on the concentration of collagen in experiments. In this section, we simply rescale the experimental $K$  by $c^{1+\delta}$ such that rescaled modulus scales as $K/c^{1+\delta} \sim c \sim \tilde{\kappa}$. Next, we obtain the individual critical strains, $\gamma_c$, for each of the concentrations as the inflection point of the $\log K$ vs.\ $\log \gamma$ curve. We then consider the experimental data (rescaled by $c^{1 + \delta}$) for each concentration along with its $\gamma_c$ and fit the entire curve to Eq.~\eqref{crossover} with $\tilde{\kappa}$ as the only free parameter. Here we show the result of the fitting for a 1mg/mL collagen network in Fig.~\ref{stiffeningcurves}(b) superimposed on the experimental data. We have reported the full set of experimental curves over a wide range of concentrations of collagen along with the fitting in Ref.~\cite{sharma2016strain}.

%As can be seen in the inset of Fig.~\ref{stiffeningcurves}b, the average value of $\tilde{\kappa}$ obtained from three samples per concentration, scales linearly with the concentration consistent with the predictions of our model. 

\begin{figure}[t]
\includegraphics[width = \columnwidth]{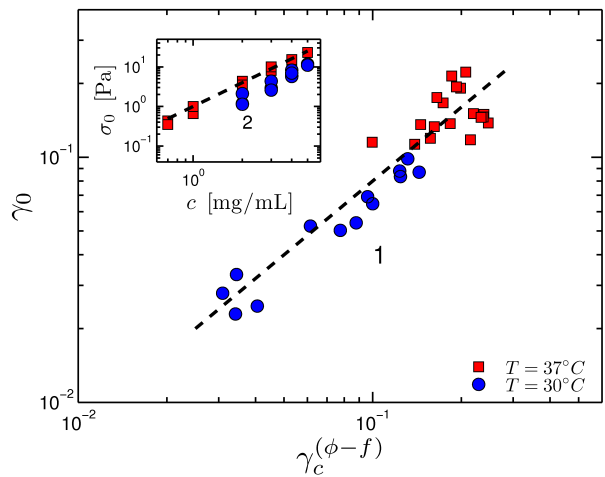}
\caption{%(a) The critical strain $\gamma_c$ depends weakly on the concentration. Since $K \sim c^2 \gamma_c^{(f-\phi)}$, the weak concentration dependence of $\gamma_c$ leads to deviation of stiffness from the expected $K \sim c^2$ (in the linear regime) scaling as shown in the inset where $K \sim c^{2.2}$. (b) 
(Color online) The onset strain for stiffening scales as $\gamma_0 \sim \gamma_c^{(\phi -f)}$. The critical exponents are $\phi = 2.1$ and $f = 0.8$. The experimental data are taken from collagen networks prepared at temperatures, $T=30^{\circ}$C ($\mathlarger{\mathlarger{\circ}}$) and $37^{\circ}$C ($\square$). This scaling is a direct consequence of the measured $c^2$ scaling of the shear stress at $\gamma_0$ as shown in the inset.}
\label{criticalstrain}
\end{figure}

\section{Relation between $\gamma_0$ and $\gamma_c$} \label{deviation}
In a recent study, we showed that the onset of stiffening strain $\gamma_0$ is practically independent of the concentration of collagen~\cite{licup2015stress}. The invariance of the geometrical structure of the network with concentration, in particular of the average connectivity in the network, was suggested as the underlying reason for the independence. The same argument leads to the conclusion that $\gamma_c$ is independent of the concentration and should be determined entirely by the geometry of the network. It is therefore expected that a general relation exists between $\gamma_0$ and $\gamma_c$.

%However, on plotting the experimentally obtained $\gamma_c$ versus concentration, a rather weak dependence on concentration is apparent as shown in Fig.~\ref{criticalstrain}. $\gamma_c$ is obtained as the inflection point of the stiffening curves in Fig.~\ref{stiffeningcurves}(b). It is plausible that with increasing concentration, the average connectivity in the network increases leading to a decrease in $\gamma_c$. The dependence of $\gamma_c$ on concentration, though weak, can in fact account for the deviation of stiffness from the $K \sim c^2$ scaling. In the limit of $\gamma \rightarrow 0$, using $\mathcal{G_{-}}(x) \sim x$ (Eq.~\eqref{collapse}), we obtain a scaling relation for the linear modulus and the concentration: $K/c \sim c \gamma_c^{(f - \phi)}$. This can account for the deviation of $K$ from the expected quadratic scaling: $K/c \sim c \gamma_c^{(f - \phi)} \sim c^{1+\delta}$ is consistent with $\gamma_c \sim c^{-0.14}$ observed in Fig.~\ref{criticalstrain}, since $\delta \simeq 0.14(\phi-f) \simeq 0.2$. The internal consistency between the $K \sim c^{2.2}$ and the concentration dependence of $\gamma_c$ provides evidence that the network connectivity evolves weakly with the concentration. 

An expression for $\gamma_0$, based on geometrical arguments has been derived in Ref.~\cite{licup2015elastic}. We can obtain an expression for $\gamma_c$ in terms of $\gamma_0$ and critical exponents in the following way. Using Eq.~\eqref{collapse}, the linear modulus $G_0$ can be written as $G_0 \equiv K(\gamma = 0) \sim c^2 \gamma_c^{f-\phi}$. It follows that the stress at the onset of stiffening should scale as $\sigma_0 = G_0 \gamma_0 \sim c^2 \gamma_c^{f-\phi} \gamma_0$. The experimentally obtained $\sigma_0$ versus concentration is shown in the inset of Fig.~\ref{criticalstrain}. The data are taken from collagen networks prepared at temperatures, $T=30^{\circ}$C and $37^{\circ}$C. As can be seen in Fig.~\ref{criticalstrain}, $\sigma_0$ scales quadratically with the concentration implying that
\begin{equation}
  \gamma_0 \sim \gamma_c^{(\phi - f)}. 
  \label{gcg0}
 \end{equation} 
  This scaling relation accurately describes the relation between $\gamma_0$ and $\gamma_c$ as shown in Fig.~\ref{criticalstrain} with $\phi = 2.1$ and $f = 0.8$. However, unlike $\gamma_0$, which can be determined analytically, determination of $\gamma_c$ from Eq.~\eqref{gcg0} requires the knowledge of the critical exponents which, at present, are only obtained from scaling analysis of stiffening data. It is important to note that the above arguments are valid only when the average connectivity in the network depends weakly on the concentration. This requirement is based on the observation, as shown in the next section, that the critical exponents evolve with the average connectivity in the network. Using a unique set of values for $\phi$ and $f$ in Eq.~\eqref{gcg0} requires that these two exponents are practically constant over the entire range of collagen concentrations.

\begin{figure}[t]
\includegraphics[width = \columnwidth]{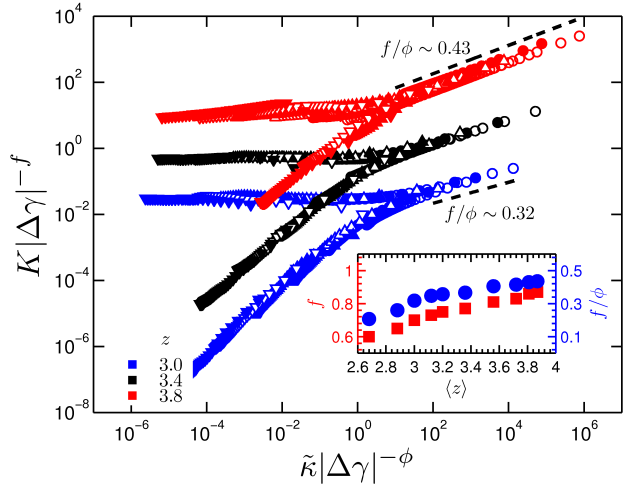}
\caption{(Color online) Shear stiffness versus shear strain curves collapsed according to the Eq.~\eqref{collapse} for phantom triangular networks in 2D prepared at different connectivities (see legend). The red and the blue data sets have been shifted by a decade up and down, respectively, for better visualization. The exponent $f$ changes significantly with $\langle z \rangle$. With $\phi$ showing practically no dependence on the connectivity, the ratio $f/\phi$ increases with the connectivity as shown in the inset.}
\label{exponents}
\end{figure}

%
%
%\begin{figure*}[ht]
%\begin{tabular}{@{}cc@{}}
%\includegraphics[width = \columnwidth]{./Figures/bulkstiffness.eps} &
%\includegraphics[width = \columnwidth]{./Figures/bulkwidom.eps} 
%\end{tabular}
%\caption{(Color online) (a) Bulk stiffness versus bulk strain obtained from a phantom triangule network in 2D with $\langle z \rangle \approx 3.2$. The network is subjected to isotropic expansion. Different curves correspond to different $\tilde{\kappa}$. (b) Data in (a) collapsed according to Eq.~\eqref{Bcollapse} shown in black. Stiffness data from a 3D FCC lattice with same connectivity as in 2D of $\langle z \rangle \approx 3.2$ is shown in red. The critical exponents, independent of the spatial dimensionality, are $f = 0.35 \pm 0.05$ and $\phi = 2.0 \pm 0.1$.  Dashed lines in (a) through the symbols are predictions of Eq.~\eqref{crossover} with $\Delta \gamma$ replaced with $\Delta \epsilon$ and the critical exponents $f = 0.35 \pm 0.05$ and $\phi = 2.0 \pm 0.1$.}
%\label{bulkstiffness}
%\end{figure*}

\section{Critical exponents and connectivity} \label{exponentsevolution}
%The critical exponents obtained by data collapsing in Fig.~\ref{widomcollapse}, for both simulations and experiments, are the same. 
Strikingly, the critical exponents obtained by collapsing both simulation data of 2D and 3D fibrous networks and experimental data for collagen networks are identical~\cite{sharma2016strain}as long as the average network connectivity is the same. The exponents are apparently independent of the spatial dimensionality. This is in contrast to both thermal and athermal critical phenomena where the critical exponents depend on the spatial dimensionality~\cite{goldenfeld1992lectures,stauffer1994introduction}. In fact, the critical exponents evolve with the average connectivity in the network. In Fig.~\ref{exponents}, we show the nonlinear stiffness data collapsed according to Eq~\eqref{collapse} for 2D triangular lattice-based networks prepared at different connectivities. The inset of Fig.~\ref{exponents} shows a plot of $f$ and $f/\phi$ versus the average connectivity for both 2D and 3D lattice-based networks. It is clear that $f$ increases with the average connectivity in the network whereas $\phi$ remains practically constant. The evolution of critical exponents with the connectivity has been also observed in branched networks modeled as diluted honeycomb structures~\cite{rens2016nonlinear}. 

The continuous variation of critical exponents is similar to the behavior of Ashkin-Teller and 8-vertex models, which exhibit continuously varying critical exponents along a critical line~\cite{ashkin1943statistics,baxter1971eight,kadanoff1979correlation}. Such a variation in the critical exponents has been experimentally observed in certain quantum phase transitions~\cite{butch2009evolution,fuchs2014critical}. In Ref.~\cite{rens2016nonlinear}, we presented a hypothesis that the apparent variation of the critical exponents could correspond to a crossover between critical exponents in the pure and disordered limits where the pure limit corresponds to an undiluted and undistorted perfect lattice based network. At present it remains unclear whether the variation can be attributed to a crossover behavior.  
%%However, it is interesting to consider the experimental verification of the the same. 
%In order to do so, the connectivity in a network needs to be systematicaly varied. %From Sec.~\ref{deviation}, we know that the average connectivity shows a very weak dependence on the concentration. Varying connectivity with concentration, though possible, is not a very practical method to verify the theoretical predictions. In an isotropic network, the total number of constraints scale (increase) with the degree of connectivity. 
However, based on previous simulations~\cite{sheinman2012nonlinear} an interesting experimental verification of varying exponents could be to isotropically compress a subisostatic random network, since this would reduce the number of constraints while leaving the connectivity the same.

\begin{figure}[t]
\includegraphics[width = \columnwidth]{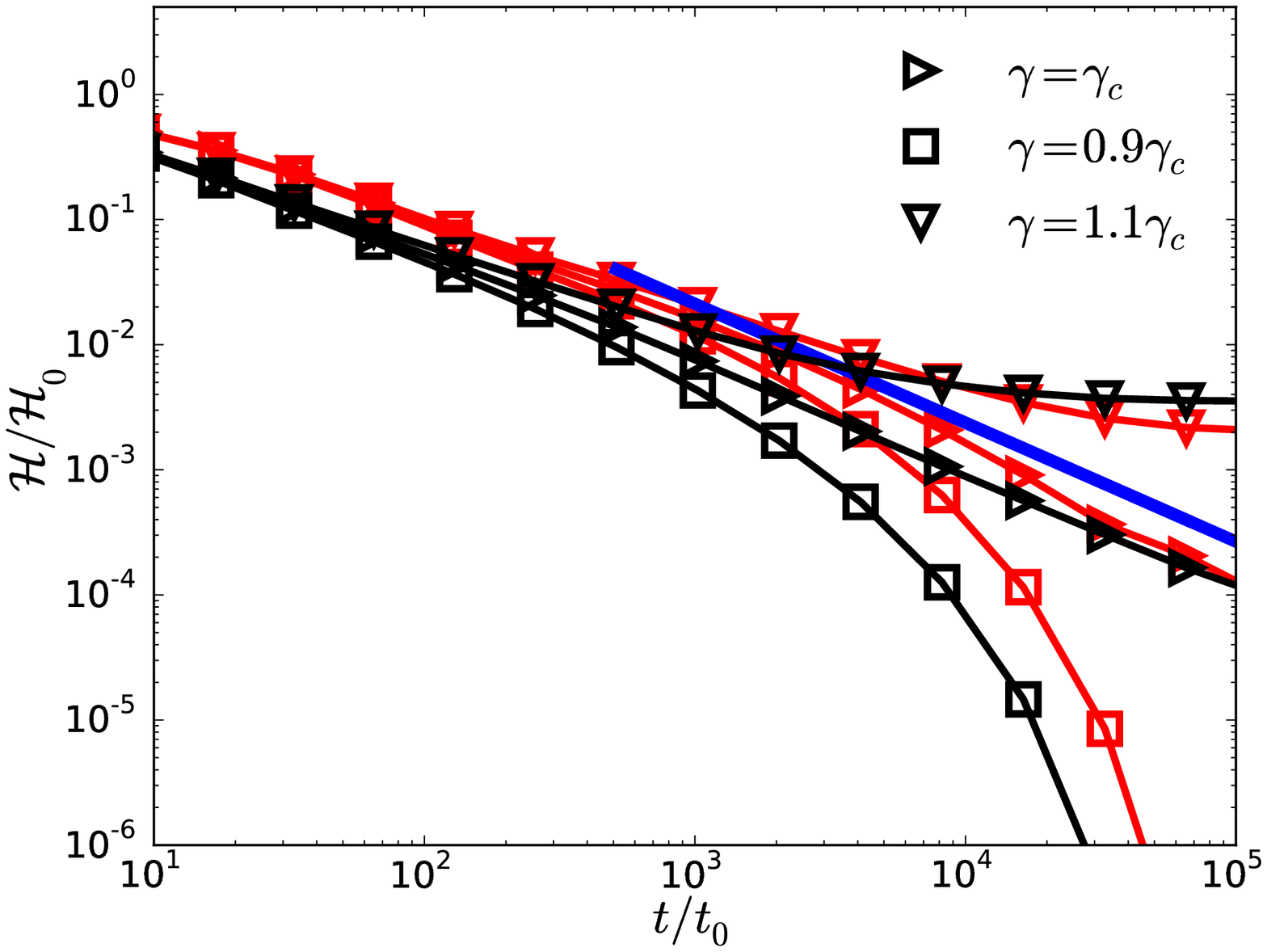}
\begin{picture}(0,0)
\put(-90,35){\includegraphics[height=2.5cm]{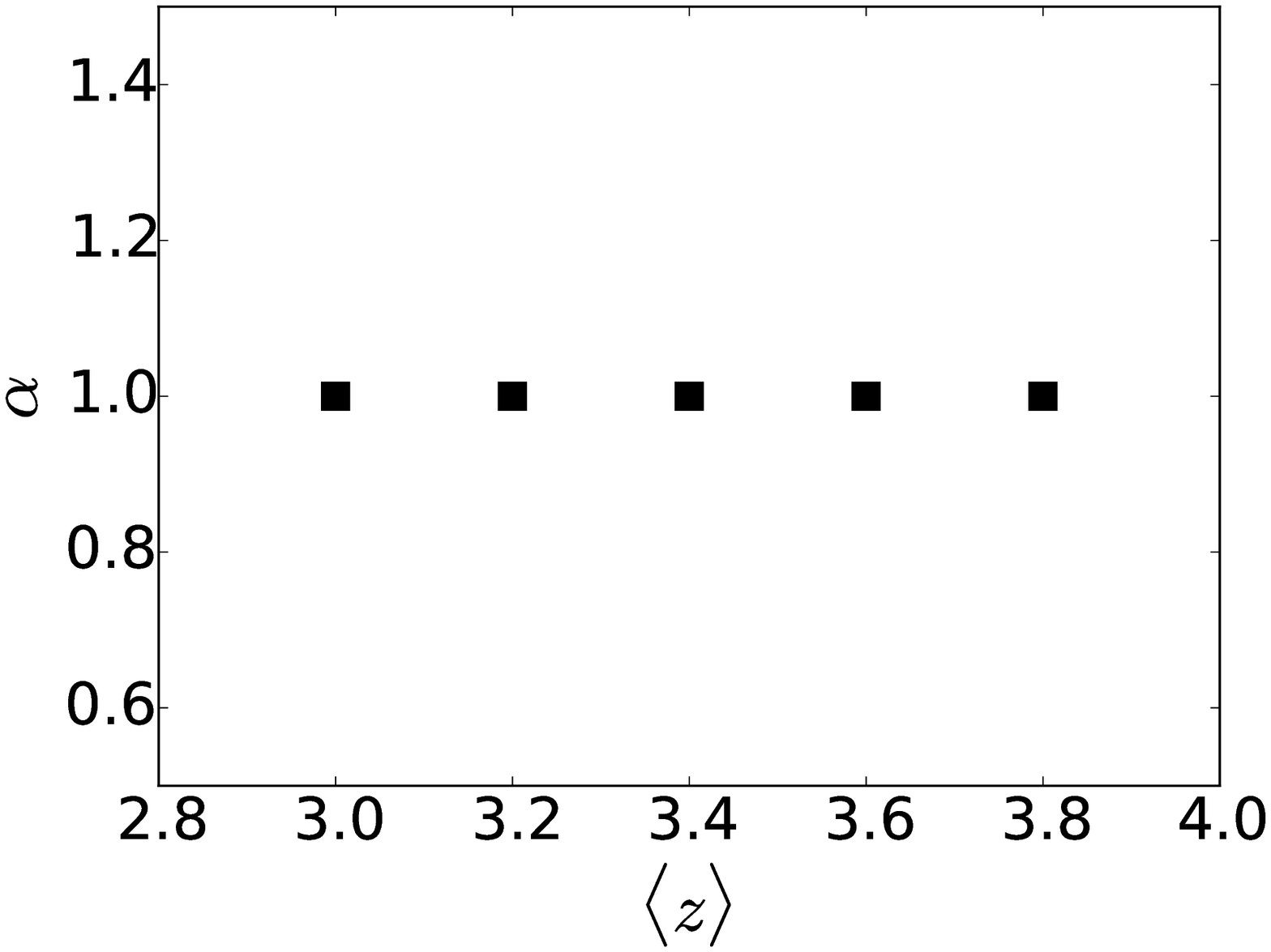}}
\end{picture}
\caption{(Color online) Energy versus time for two different network connectivities, $\langle z \rangle \simeq 3.6$ (red) and $\langle z \rangle \simeq 3.2$ (black) for simple shear. The bending rigidity $\tilde{\kappa}$ is set to 0 to take only the central-force interactions into account. Energy is expressed in units of the initial energy in the network just after the affine deformation, $\mathcal{H}_0$. The time is expressed in arbitrary units, chosen to be the same for both connectivities. For both connectivities, elastic energy stored in the network decays as a function of time. Very close to the critical strain, the relaxation dynamics follow a power-law $E(t) \sim t^{-1}$ indicated with the thick blue line. The exponent $\alpha$ for critical slowing down is insensitive to the network connectivity in the range of 3.0-3.8 as shown in the inset.}
\label{bulkslowingdown}
\end{figure}

\section{Critical slowing down} \label{slowdynamics}
One of the hallmark signatures of a critical phenomenon is extremely slow dynamics at the critical point~\cite{goldenfeld1992lectures}. The dynamics are characterized by a divergent relaxation time scale. In a disordered fibrous network, we investigate the critical slowing down by applying an affine deformation to the network such that the strain equals the critical value. We only take central-force interactions into account by setting $\tilde{\kappa}=0$. We then let the network relax the elastic energy by performing overdamped Molecular Dynamics simulations. We do not take hydrodynamics into account. We also ignore the asymmetric nature of drag acting on each filament. We rather assume that the drag forces acting on the network due to the surrounding solvent can be modeled in a simple Stokesian fashion and can be lumped on the network node. This is admittedly a highly simplified version of network dynamics. We subjected central-force subisostatic networks with connectivities in range of $3.0-3.8$ to an affine shear of $\gamma = \gamma_c, 0.9\gamma_c$ and $1.1\gamma_c$. The network is floppy for $\gamma \leq \gamma_c$ implying that the total elastic energy stored in the network decays to zero in the long-time limit. Since the network is rigid for $\gamma > \gamma_c$, the total elastic energy should relax to a finite value after a characteristic relaxation time. In Fig.~\ref{bulkslowingdown}, we show the time evolution of the total elastic energy stored in the network for two connectivities $\langle z \rangle = 3.2$ and $3.6$. Clearly, for $\gamma \lessgtr \gamma_c$, there is a characteristic relaxation time. However, at $\gamma = \gamma_c$ the slowed down dynamics are robustly captured in the power law scaling of the total elastic energy in the network as a function of time. For longer times, the elastic energy stored in the network decays as $E(t) \sim t^{-\alpha}$ at $\gamma = \gamma_c$ with $\alpha \simeq 1$ implying that the relaxation time scale is divergent. This inverse-time decay is apparent in all the connectivities considered in this study. Unlike the critical exponents $f$ and $\phi$, the exponent associated with critical slowing down does not evolve with connectivity.

 % For strains above and below the critical strain, the relaxation time scale is finite. 
The divergent time scale of relaxation at the critical point has its origin in the highly delocalized structural rearrangements in the network. These structural rearrangements are deviations from the imposed affine deformation and are apparent as divergent strain fluctuations as shown in Fig.~\ref{diffNA}. In the thermodynamic limit of $W \rightarrow \infty$, the non-affine rearrangements in the network grow without bound giving rise to the divergent time scale of energy relaxation.

\section{Discussion and conclusions} \label{conclusions}
In this study, we focus on the mechanical critical behavior in fiber networks. The networks considered are athermal, disordered, and are by construction, subisostatic. The criticality is driven by the applied global deformation and is the fundamental mechanism of the nonlinear mechanics of such networks. Unlike the isostatic connectivity threshold which depends on the precise balance of the number of constraints to the degrees of freedom, any generic subisostatic network exhibits critical behavior when subjected to an external deformation. The criticality is evident in the neighborhood of a strain that is determined by the network architecture.

%We take into account two types of interactions: central force as in a Hookean spring and bending forces. 

 % However, not all types of deformations drive critical behavior, thus preventing generalisation of criticality when considering an arbitrary deformation as in a strain tensor. This is easily seen in a subisostatic network with central force interactions, subjected to isotropic compression.

One of the hallmark features of critical phenomena is the power-law scaling of the order parameter in the vicinity of the critical point. We show that the stiffness of subisostatic networks with central-force interactions scales as a power-law, $\mathcal{K} \sim |\Delta \gamma|^f$, where $\Delta \gamma = \gamma - \gamma_c \geq 0$ is the distance measured from the critical strain and $f$ is a critical exponent. Additional interactions such as resistance to bending stabilize subisostatic networks in the subcritical regime $\Delta \gamma < 0$ such that for $\gamma << \gamma_c$, $\mathcal{K} \sim \tilde{\kappa}$ where $\tilde{\kappa}$ is the bending rigidity. From the perspective of a critical phenomenon, finite bending rigidity can be considered as an auxiliary field that suppresses the strain-driven criticality. For $\tilde{\kappa} > 0$ the stiffness at the critical strain is finite and depends in a power-law fashion on the strength of bending and stretching interactions. Drawing analogy with the ferromagnetic phase transition, where $H$, the applied magnetic field is the auxiliary field, we capture the crossover of stiffness from bend-dominated to stretch-dominated regimes in terms of a universal scaling function.

Another important signature of criticality besides the power-law scaling of the order parameter is the divergence of fluctuations in the order parameter at the critical point. In athermal subisostatic networks, the order parameter $\mathcal{K}$ is zero at the critical strain and exhibits no fluctuations. However, on considering the deviation of the strain field within the network from the globally imposed affine field, one can create measures for fluctuations. We construct one such measure: the strain-derivative of average bending-angle in the network and using finite size scaling demonstrate its divergence in the thermodynamic limit. Recently Xu \emph{et. al} have developed an image analysis software SOAX which can accurately track fibers in 3D~\cite{xu2015soax}. It is an interesting idea to use SOAX together with confocal shear cell rheology~\cite{arevalo2015stress} to experimentally measure the average bending angle in reconstituted biopolymer networks. 

We also study a highly simplified model of network dynamics to test if the network relaxation at the critical point exhibits signatures of critical slowing down. We subject subisostatic networks to an affine shear and study the relaxation of the total elastic energy in the network as a function of time. We find that the elastic energy decays as a power-law in time as $\sim t^{-1}$ at the critical strain. The power-law decay implies a divergent relaxation time at the critical strain. We find that the dynamics of networks prepared over a wide range of connectivity $\langle z \rangle = 3.0-3.8$ remain the same, i.e., the critical exponent associated with slowing down at the critical strain appears to be insensitive to the connectivity in the network.

 The analogy with the ferromagnetic phase transition guides us in writing an approximate equation for the scaling function that captures the crossover of stiffness from bend-dominated to stretch-dominated regimes. We demonstrate that the derived equation is highly accurate in describing the entire nonlinear stiffness vs. strain curves for any bending rigidity. Since concentration in experiments can be mapped to the reduced bending rigidity in our network model, the equation for the crossover function can equivalently describe the stiffness vs. strain curves for any concentration of the protein in the experiments. We show that the equation accurately describes the stiffness of collagen networks with a single fit parameter. The excellent agreement of model predictions with the experiments provides strong evidence for criticality as the underlying mechanism of the well known nonlinear mechanics of athermal fibrous networks such as collagen~\cite{sharma2016strain, licup2015stress, motte2013strain,piechocka2011rheology} and bundled actin~\cite{Weitz2004,Janmey2005,Kasza2009}.

A surprising observation is that under simple shear, the critical exponents $f$ and $\phi$ appear to be independent of the spatial dimensionality. %Networks with same average connectivity have the same critical exponents independent of the dimensionality. 
This is a highly intriguing and also puzzling observation. The critical exponents, as is known from the theory of critical phenomena, depend on the spatial dimensionality. However, the exponents are not constant as they change with the average connectivity in the network. %The exponent $f$ increases with the connectivity and $\phi$ remains practically constant. 
The variation of critical exponents along a critical line is similar to the Ashkin-Teller and 8-vertex models~\cite{ashkin1943statistics,baxter1971eight,kadanoff1979correlation}.

The variation in the exponents occurs over a range of connectivities that is significantly larger than that found in collagen networks. Therefore, one can use a unique set of exponents, $f \simeq 0.8$ and $\phi \simeq2.1$ to describe the mechanics of collagen networks prepared at different concentrations~\cite{sharma2016strain}. The uniqueness of the exponents also allows us to relate the two characteristic strains of a subisostatic network, onset of stiffening strain and critical strain via the critical exponents as $\gamma_0 \sim \gamma_c^{\phi - f}$.
%It becomes even more interesting when considering the critical exponents of a network subjected to isotropic expansion. Under isotropic expansion, just as in shear, the critical exponents do not depend on the spatial dimensionality. However, unlike shear, the exponents appear to be constant. The same analytical approximation used to predict the stiffness curves in simple shear can accurately predict the stiffness in isotropic expansion. 

%We also study a highly simplified model of network dynamics to test if the network relaxation at the critical point exhibits signatures of critical slowing down. We find that the elastic energy decays as a power-law in time as $\sim t^{-1}$ at the critical strain. The power-law decay implies a divergent relaxation time at the critical strain. For strains above and below the critical strain, the The dynamics remain the same on varying the connectivity. 

In sum, the mechanics of disordered fibrous networks can be understood within the framework of an athermal strain-driven critical phenomenon. The mechanical criticality is a generic phenomenon exhibited by all subisostatic networks. We apply our model to collagen networks which are ubiquitous in biology and find strong evidence for the idea that mechanical critical behavior underlies the strain-stiffening response of collagenous networks
%Mechanical critical behavior has been studied in the past in networks with connectivity satisfying the isostatic threshold of $z = 2d$. Naturally occuring networks are almost always sub-iosostatic. 
%
%
%In contrast to most prior work on critical phenomena for jamming \cite{van2010jamming}, rigidity percolation \cite{thorpe1983continuous,feng1984percolation} and fiber networks \cite{wyart2008elasticity,chase2011} near isostaticity, our focus here has been on networks well below the isostatic point. This is most relevant to networks such as those of collagen, especially in 3D. One challenge in understanding such systems has been their nonlinear mechanical response. Recently, a Landau type theory for the non-linear elasticity of biopolymer gels was proposed using an order parameter describing induced nematic order of fibers in the gel\cite{feng2014alignment}. Our model here can provide a framework to understand the nonlinear mechanics of extracellular networks such as collagen in terms of critical phenomena associated with the isostatic point. Importantly, as we show, there is actually a line of critical points that extends over a wide range of network connectivities, covering the physiologically relevant range of $z=3-4$ in 3D. Moreover, although the linear modulus of collagen networks may be finite in this range due to the stabilizing influence of bending, the nonlinear response can be quantitatively captured by the scaling functions in Eqs.\ (\ref{collapse},\ref{exptcollapse}).

\bibliographystyle{apsrev}
\bibliography{references_new}

\clearpage

\appendix

\end{document}